\documentclass[%
 reprint,
superscriptaddress,
 amsmath,amssymb,
 aps,
prb,
]{revtex4-2}

\usepackage[dvipdfmx]{graphicx}% Include figure files
\usepackage{dcolumn}% Align table columns on decimal point
\usepackage{bm}% bold math
\usepackage{color}
\begin{document}

%%\preprint{APS/123-QED}

\title{Antiferromagnetic order of ferromagnetically coupled dimers in the double pyrovanadate CaCoV$_2$O$_7$}% Force line breaks with \\
%%\thanks{A footnote to the article title}%

\author{Ryo Murasaki}
\email{r.murasaki@dc.tohoku.ac.jp}
\author{Kazuhiro Nawa}
\author{Daisuke Okuyama}

\affiliation{%
  Institute of Multidisciplinary Research for Advanced Materials, Tohoku University,
  2-1-1 Katahira, Sendai 980-8577, Japan
}%

\author{Maxim Avdeev}
\affiliation{Australian Centre for Neutron Scattering, Australian Nuclear Science and Technology Organisation,
  Locked Bag 2001, Kirrawee, NSW 2232, Australia}
\affiliation{School of Chemistry, The University of Sydney, Sydney, NSW 2006, Australia}

\author{Taku J Sato}
\email{taku@tohoku.ac.jp}
\affiliation{%
  Institute of Multidisciplinary Research for Advanced Materials, Tohoku University,
  2-1-1 Katahira, Sendai 980-8577, Japan
}%

\date{\today}% It is always \today, today,
             %  but any date may be explicitly specified

\begin{abstract}
  Magnetic properties of the pyrovanadate CaCoV$_2$O$_7$ have been studied by means of the bulk magnetization and neutron powder diffraction measurements.
  Magnetic susceptibility in the paramagnetic phase shows Curie-Weiss behavior with negative Weiss temperature $\Theta \simeq -22.5$~K, indicating dominant antiferromagnetic interactions.
  At $T_{\rm N} = 3.44$~K, CaCoV$_2$O$_7$ shows antiferromagnetic order, accompanied by a weak net ferromagnetic moment of $\sim 0.05$~$\mu_{\rm B}/{\rm Co}^{2+}$.
  Neutron powder diffraction confirms the formation of antiferromagnetic order below $T_{\rm N}$.
  It was further confirmed from the magnetic structure determination that the two Co$^{2+}$ ions in the adjacent edge-sharing octahedra have almost parallel (ferromagnetic) spin arrangement, indicative of a formation of a ferromagnetic spin dimer.
  The antiferromagnetic order is, in turn, stabilized by sizable inter-dimer antiferromagnetic interactions.
\end{abstract}

%\keywords{Suggested keywords}%Use showkeys class option if keyword
                              %display desired
\maketitle

%\tableofcontents

\section{Introduction}

Structural~\cite{MercurioLavaud1973,Calvo1975,Liao1996,Touaiher2004,KrivovichevSV2005,HeZ2008}, magnetic~\cite{Liao1996,Touaiher2004,TsirlinAA2010,SannigrahiJ2017,BhowalS2017,JiWH2019,YinL2019}, magnetoelectric~\cite{Sannigrahi2015,LeeYW2016,ChenR2018,ChenR2019,WuHC2020}, and photocatalytic~\cite{GuoW2015,YanQ2015,KimMW2017,CamargoLP2020,SongA2020} properties of the transition-metal pyrovanadates with the chemical formula $M_2$V$_2$O$_7$ ($M =$ Mn, Co, Ni and Cu) attract continuous interest for decades.
Building blocks of the pyrovanadates are typically pairs of corner-shearing nonmagnetic VO$_4$ tetrahedra, forming a so-called pyroanion [V$_2$O$_7$]$^{4-}$.
The pyroanion connects the transition metal $M^{2+}$ magnetic ions in various manner, realizing a wide variety of magnetic lattice, ranging from one-dimensional chains to very complex three-dimensional networks.
A representative example may be $\alpha$-Cu$_2$V$_2$O$_7$, where the magnetic Cu$^{2+}$ ions form a three-dimensional helical-honeycomb structure~\cite{CalvoC1975,RobinsonPD1987,Gitgeatpong2015}.
Originating from the noncentrosymmetric crystal structure of $\alpha$-Cu$_2$V$_2$O$_7$, an enhancement of the electric polarization was observed upon the magnetic ordering, while intriguing Rashba-type magnon splitting was reported in the ordered phase~\cite{LeeYW2016,BanerjeeA2016,ChattopadhyayB2017,GitgeatpongG2017,BhowalS2017,ShiomiY2017,ZhangJT2017,GitgeatpongG2017b,WangL2018}.
A formation of a distorted honeycomb lattice is also seen in the Mn$_2$V$_2$O$_7$ pyrovanadate, where intriguing coexistence of ferroelastic and antiferromagnetic ordering was reported~\cite{SannigrahiJ2019,WuHC2020}.
On the other hand, bond-alternating skew chains form in Co$_2$V$_2$O$_7$, in which the magnetic-field-induced magnetization plateau, possibly related to the ferroelectricity, was observed~\cite{SannigrahiJ2017,ChenR2018,YinL2019,ChenR2019,JiWH2019}.

\begin{figure}
  \includegraphics[scale=0.3, angle=90, trim=0 0 0 0]{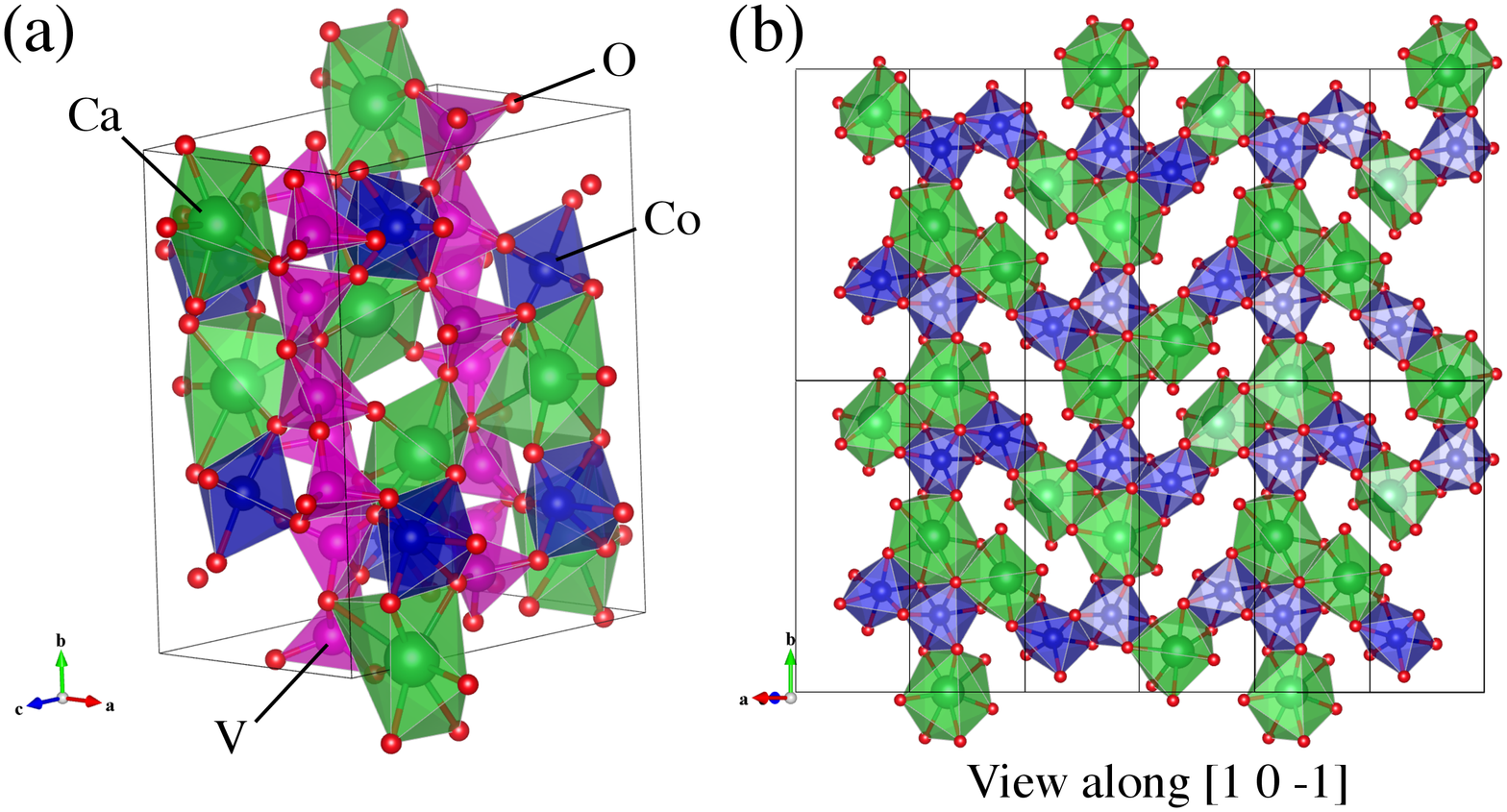}
  \caption{\label{figure0} (a) Crystal structure of CaCoV$_2$O$_7$~\cite{MurashovaEV1993}.
    (b) A 2D network of edge-sharing Co and Ca octahedra, formed in a plane normal to the crystallographic $[10\bar{1}]$ direction.
    This Co-Ca layer is sandwiched by the vanadium-tetrahedra layers.
    Edge-sharing pairs of Co octahedra can be clearly seen in the structure.
    Structure illustrations were made with VESTA~\cite{MommaK2011}.
  }
\end{figure}

Closely related to the pyrovanadates $M_2$V$_2$O$_7$, there exists a class of materials known as the double pyrovanadates with the chemical formula $M$$M'$V$_2$O$_7$, where $M$ and $M'$ are typically (but not limited to) the nonmagnetic alkaline-earth and magnetic transition-metal ions, respectively.
Crystal structure of the double pyrovanadates has been extensively studied in the past~\cite{ZhuravlevV2018,BabarykAA2015}.
It has been known that introduction of nonmagnetic $M$ ions results in deformation and/or termination of magnetic lattice in various manner.
For instance, in CaCuV$_2$O$_7$, the nonmagnetic Ca$^{2+}$ ions replace the magnetic Cu$^{2+}$ ions alternatingly, resulting in a formation of rather magnetically isolated CuO$_6$ octahedra~\cite{Vogt1991}.
In contrast, Ca substitution works quite differently in CaCoV$_2$O$_7$, where a pair of edge-sharing CoO$_6$ octahedra is found in the crystal structure, as shown in Fig.~\ref{figure0}~\cite{MurashovaEV1993}.
The edge-sharing CoO$_6$ octahedra are linked by [V$_4$O$_{14}$]$^{8-}$ tetramers, which are composed of a pair of pyroanions [V$_2$O$_7$]$^{4-}$.
Tailoring crystal and magnetic lattices may thus be possible in the double pyrovanadates, making this class of material a very attractive playground for searching interesting magnetic and magnetoelectric properties.

To date, however, relatively little has been explored for the magnetic properties of the double pyrovanadates.
Therefore, in the present work we undertook a bulk-magnetic, X-ray and neutron diffraction characterization of the representative double pyrovanadate compound CaCoV$_2$O$_7$.
The powder X-ray and neutron diffraction both confirm the reported crystal structure with the $P2_1/c$ space group.
The dominant magnetic interaction is found to be antiferromagnetic from Curie-Weiss fitting of the magnetic susceptibility in the paramagnetic phase.
With lowing temperature, the antiferromagnetic order, accompanied by the weak ferromagnetism, was detected as an anomaly in the bulk magnetic susceptibility.
The magnetic transition temperature was found to be $T_{\rm N} = 3.44$~K in the neutron powder diffraction experiment.
Neutron powder diffraction further confirms the dominant antiferromagnetic order below $T_{\rm N}$, in which intriguing ferromagnetic arrangement of the two adjacent Co1 and Co2 spins was observed, indicating a formation of the ferromagnetic spin dimers.

\section{Experimental}
Powder samples of CaCoV$_2$O$_7$ were prepared using the solid-state reaction method from the starting chemicals CaCO$_3$, Co$_2$O$_3$, and V$_2$O$_5$, with the typical purity being 99.99~\%, 99.9~\%, and 99.99~\%, respectively.
Several batches of the powder samples were prepared under mostly the same condition; the solid-state reaction was repeated a few times at different temperatures ranging from 550$^{\circ}$C to 750$^{\circ}$C.
Phase purity of the resulting powder sample was checked using X-ray powder diffraction with the CuK$\alpha$ radiation (Rigaku, RINT-2200).

Bulk magnetic properties were measured using the superconducting quantum interference device (SQUID) magnetometer (Quantum Design, MPMS-XL5).
The measurement was performed in the temperature range of $2 < T < 350$~K, and in the magnetic field range of $-5 \leq \mu_0 H \leq 5$~T.

The neutron powder diffraction experiment has been performed using the high-resolution powder diffractometer ECHIDNA installed at the OPAL reactor, Australian Nuclear Science and Technology Organisation.\cite{Avdeev2018}
Neutrons with $\lambda = 2.4395$~\AA\ were selected using the Ge 331 reflections.
The sample was loaded in the vanadium sample can, and then was set to the $^4$He cryostat with the base temperature 1.6~K.
Obtained powder diffraction patterns were analyzed using the Rietveld method combined with magnetic representations analysis, performed using the {\sf Fullprof} suite~\cite{fullprof1993}, together with the home-made magnetic structure analysis code~\cite{MSAS2019}.

Temperature dependence of the magnetic diffraction intensity has been measured using the thermal-neutron triple-axis spectrometer 4G-GPTAS installed at the JRR-3 reactor, Tokai, Japan.
The spectrometer was operated in the double-axis mode without using the analyzer.
Pyrolytic graphite (PG) 002 reflections were used for the monochromator with $E_{\rm i} = 14.7$~meV, and the horizontal collimations of 40$'$-40$'$-40$'$ were employed.
Powder sample of approximately 10 grams was loaded in the Al sample can with the He exchange gas, and was cooled down to 2.4~K using a closed cycle $^4$He refrigerator.

\section{Experimental results and discussion}

\subsection{X-ray characterization}

\begin{figure}
  \includegraphics[scale=0.35, angle=-90]{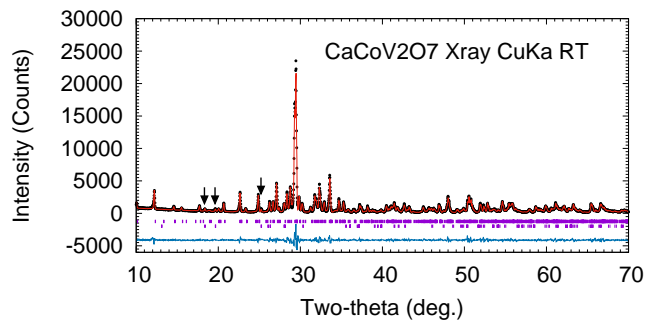} 
  \caption{\label{xraydiffraction} Room temperature X-ray powder diffraction pattern obtained from the presently prepared sample.
    Rietveld fitting results are shown by the solid lines.
    Vertical lines indicate Bragg reflection positions for CaCoV$_2$O$_7$ (upper) and $\alpha$-CaV$_2$O$_6$ (lower).
    Vertical arrows stand for the visible impurity Bragg reflections from the $\alpha$-CaV$_2$O$_6$ phase.
  }
\end{figure}

Figure~\ref{xraydiffraction} shows the X-ray powder diffraction pattern from the CaCoV$_2$O$_7$ powder sample measured at room temperature.
Also shown by the vertical bars are the positions of Bragg reflections calculated from the reported crystal structure of CaCoV$_2$O$_7$~\cite{MurashovaEV1993}.
The observed peak positions are mostly consistent with the calculated ones, confirming the quality of the presently prepared sample.
It should be noted that a closer inspection indicate an appearance of a few very weak peaks from the impurity $\alpha$-CaV$_2$O$_6$ phase (marked by the vertical arrows in the figure).
To estimate the volume fraction of the impurity phase, Rietveld analysis was performed assuming the main CaCoV$_2$O$_7$ and the impurity $\alpha$-CaV$_2$O$_6$ phases.
Initial parameters are assumed to be those reported in Ref.~\cite{MurashovaEV1993} and Ref.~\cite{BoulouxJC1972}.
The fitting result is also shown in Fig.~\ref{xraydiffraction}, together with the difference between the calculated and experimental intensities at the bottom of the figure.
From the analysis, we estimate that the volume fraction of the secondary $\alpha$-CaV$_2$O$_6$ phase is approximately 5~\%; we believe that this amount of the secondary phase will not affect the result of the present study.

\subsection{Bulk magnetic measurements}

\begin{figure}[h]
  \includegraphics[scale=0.28, angle=-90]{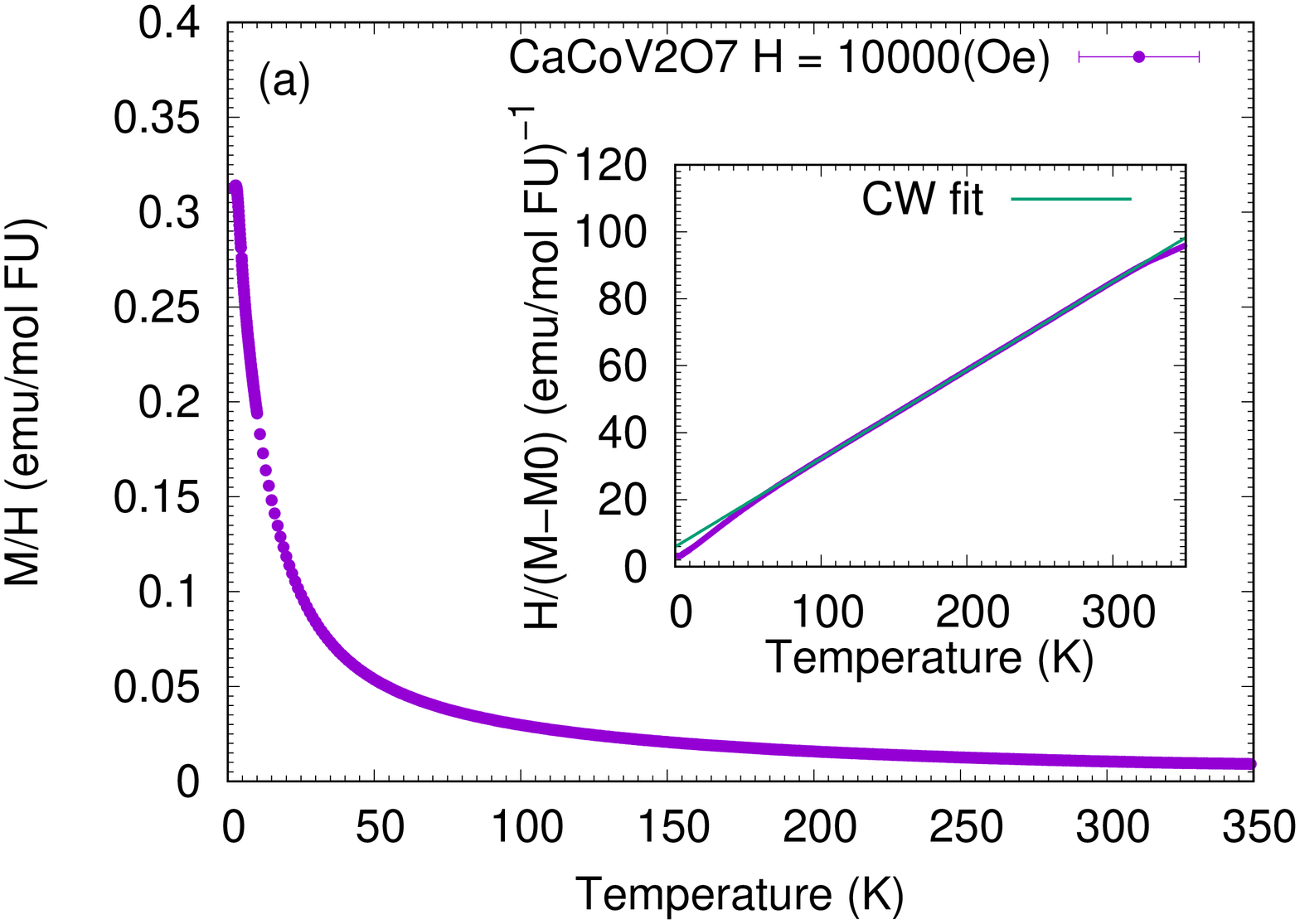}
  \includegraphics[scale=0.28, angle=-90]{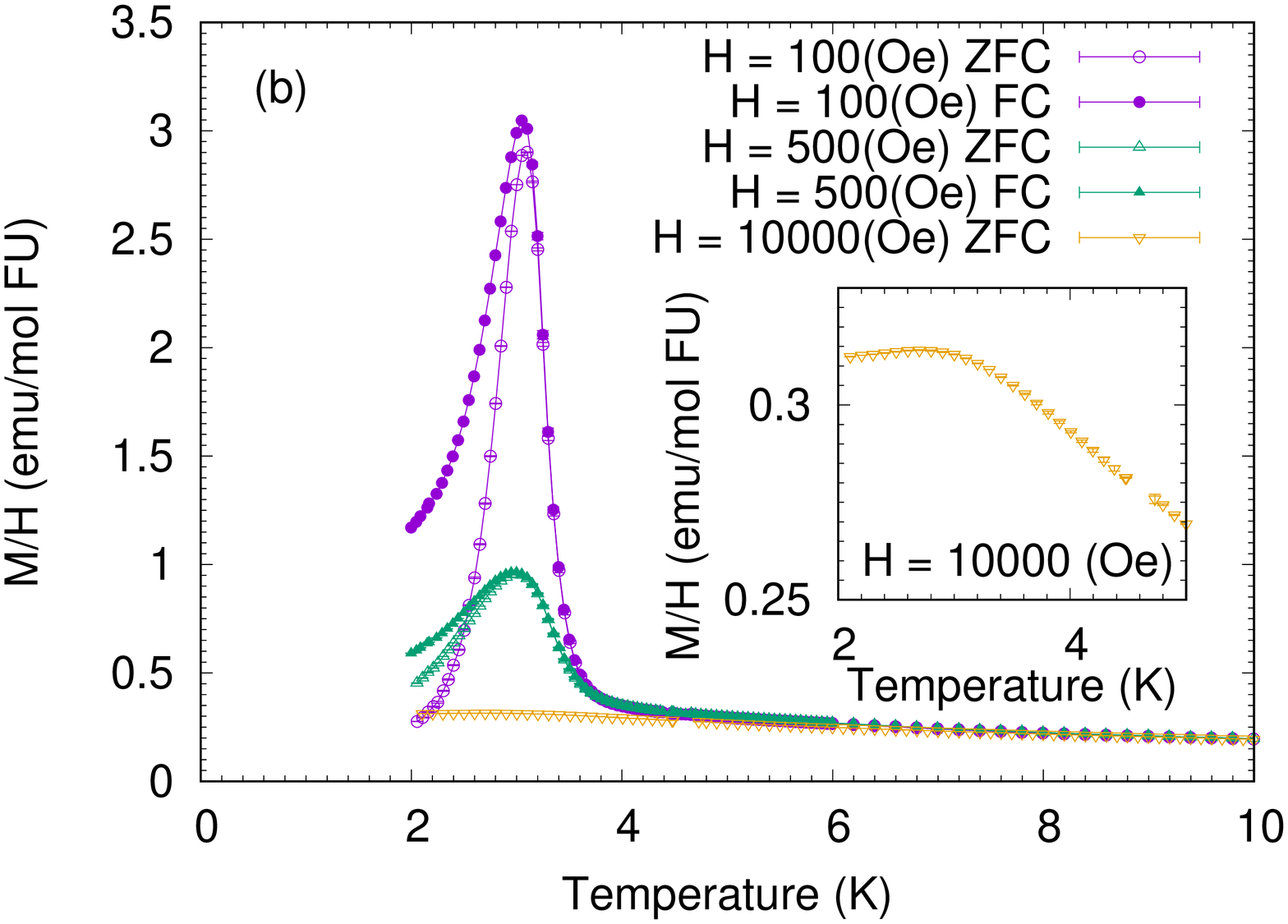}
  \includegraphics[scale=0.28, angle=-90]{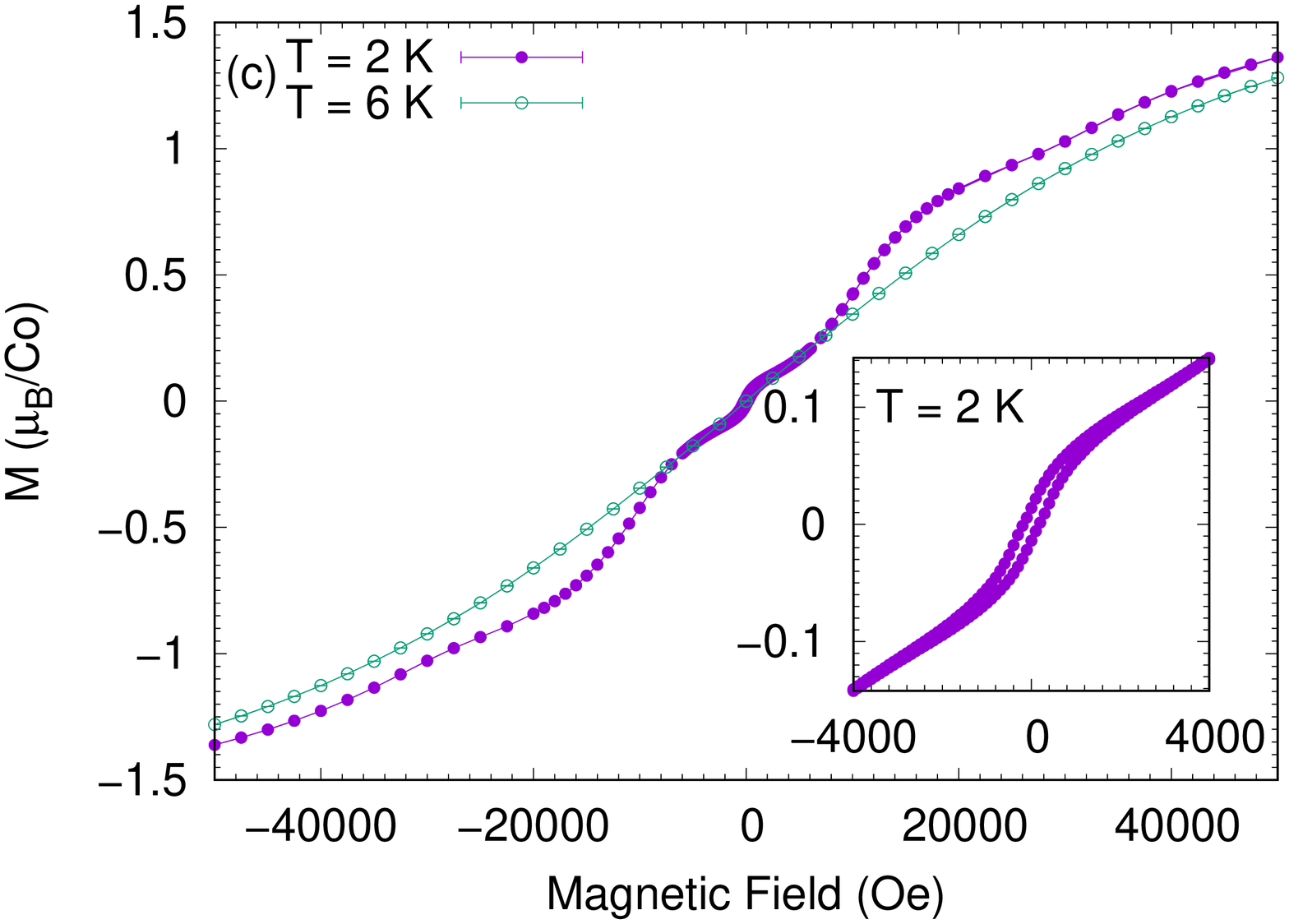}
  \caption{\label{bulkmagneticmeasurement} (a) Temperature dependence of magnetic susceptibility ($M/H$) of CaCoV$_2$O$_7$ measured in a wide temperature range $2 < T < 350$~K under the external magnetic field of $H = 10$~kOe.
    Inset: inverse susceptibility ($H/(M - M0)$).
    (b) Temperature dependence of $M/H$ in the low temperature range $T < 10$~K measured under the three external magnetic fields $H = 100$, 500 and 10000~Oe.
    Both field cooling (FC) and zero-field cooling (ZFC) results are shown for $H = 100$ and 500~Oe data.
    (c) Magnetization curves measured at $T = 2$ and 6~K, as representative results for the ordered and paramagnetic phases, respectively.
    Inset: magnified plot for the low-field region of the $T = 2$~K data.
  }
\end{figure}

Magnetic susceptibility in a wide temperature range $2 < T < 350$~K was first measured under the external magnetic field $H = 10$~kOe.
The resulting temperature dependence of $M/H$ is shown in Fig.~\ref{bulkmagneticmeasurement}(a), which shows clear increasing behavior as the temperature is lowered.
Except for the temperature range near the magnetic transition, we found that the magnetic susceptibility $\chi$ is reasonably approximated by $M/H$ at small $H$.
The observed data were fitted to the Curie-Weiss law $\chi = C/(T - \Theta) + \chi_0$, where $C$ and $\Theta$ are the Curie constant ($C = N g^2 \mu_{\rm eff}^2/3k_{\rm B}$) and Weiss temperature, respectively.
The temperature independent susceptibility $\chi_0$ was included to account for the Van Vleck and diamagnetic susceptibilities.
In the fitting procedure, we found that the Curie-Weiss fitting was best performed in a temperature range $100 < T < 300$~K; a clear deviation from the Curie-Weiss law can be seen below 100~K, plausibly originating from the splitting of the crystalline-electric-field (CEF) ground state for the $3d^7$ electrons due to spin-orbit coupling, whereas deviation of unknown origin was seen above 300~K.
The inverse susceptibility, together with the result of Curie-Weiss fitting, is shown in the inset of Fig.~\ref{bulkmagneticmeasurement}(a).
Satisfactorily linear behavior was confirmed in the temperature range.
The estimated parameters are $\mu_{\rm eff, 100-300~K} \simeq 5.5~\mu_{\rm B}$ and $\Theta \simeq -22.5$~K.
It may be noteworthy that the estimated effective moment size is close to the theoretical value of completely unquenched Co$^{2+}$.
Nonetheless, a care should be given for the quantitative validity of the estimated parameters, since the precise estimation of these parameters is intricate as it still weakly depends on the fitting temperature range.
At minimum, from the Curie-Weiss analysis we conclude a significant orbital contribution for the present Co$^{2+}$ ions, as well as dominant antiferromagnetic interactions between them in CaCoV$_2$O$_7$.

At further lower temperatures, we find clear anomaly in the temperature dependence of the magnetization.
Figure~\ref{bulkmagneticmeasurement}(b) shows the low temperature result ($T < 10$~K) measured under the external field of $H = 100$, 500, and 10000 Oe.
At the low fields, a sharp increase of the magnetization was observed at $T \simeq 3.5$~K, indicating an appearance of a long-range ferromagnetic component at this temperature.
The magnetization peaks slightly below the anomaly temperature $T_{\rm peak} \simeq 3$~K, and then decreases drastically as temperature is lowered.
It should be noted that below $T_{\rm peak}$ the bifurcation of the zero-field cooling (ZFC) and field-cooling (FC) runs can be clearly seen.
At $H = 10$~kOe, the peak in $M/H$ is largely suppressed, and only slight decrease may be seen below $T < 3$~K.

Magnetic field dependence of the magnetization ($M$-$H$ curve) is measured at the base temperature 2~K and the paramagnetic temperature $T = 6$~K up to 50~kOe.
The resulting $M$-$H$ curves are shown in Fig.~\ref{bulkmagneticmeasurement}(c), whereas the inset is a magnified plot for the low-field region.
The magnetization measured at 6~K shows smooth behavior as a function of $H$ with weak saturation at higher fields, being a typical paramagnetic behavior.
In contrast, a hysteresis can be seen in the $M$-$H$ curve measured at 2~K in the low-field region.
The size of the remnant magnetization is $\sim 0.05~\mu_{\rm B}$/Co$^{2+}$, which is very small compared to the expected moment size for Co$^{2+}$ ions.
Above the coercive field $H \simeq 1$~kOe, the magnetization shows linear increase up to the next anomaly at $\sim 10$~kOe.
In view of the small size of the ferromagnetic component ($\sim 0.05~\mu_{\rm B}$), the decreasing behavior of $\chi(T)$ below $T_{\rm peak}$, and the linearly increasing behavior of $M(H)$ above the coercive field 1~kOe, it is most likely that an antiferromagnetic ordering accompanied by a weak ferromagnetic component is realized below $T = 3.5$~K.
We note that metamagnetic transitions were observed at the higher fields $H \simeq 10$ and 30~kOe in the $M$-$H$ curve at $T = 2$~K.
Since it is generally hard to identify the microscopic origin of the metamagnetic transitions solely from the powder magnetization data, we will not go into details, but just note here that such metamagnetic transition is not unexpected for magnets with Co$^{2+}$ ions which generally have strong single-ion anisotropy.

\subsection{Neutron diffraction}
\begin{figure*}
  \includegraphics[scale=0.3, angle=-90]{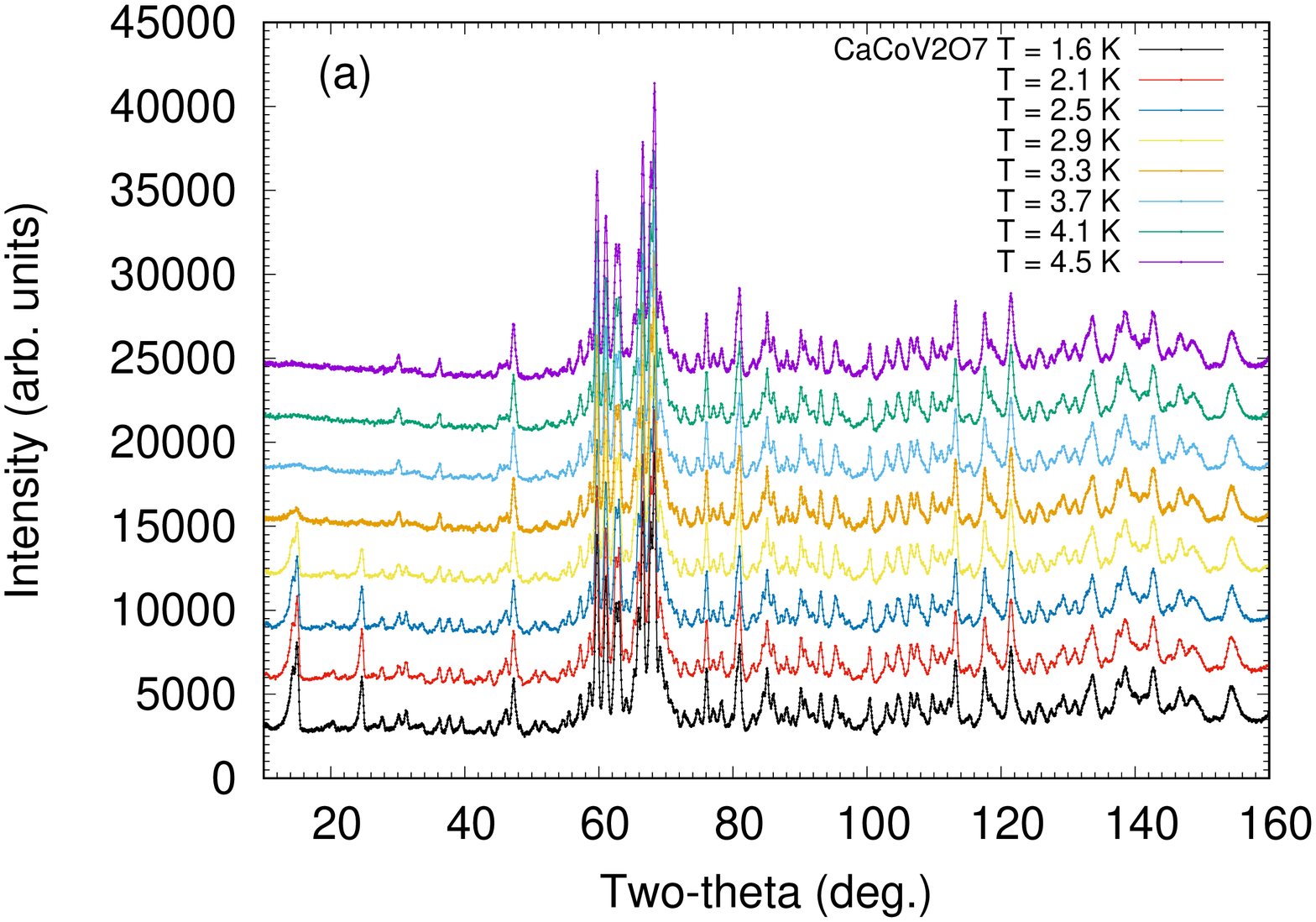}
  \includegraphics[scale=0.3, angle=-90]{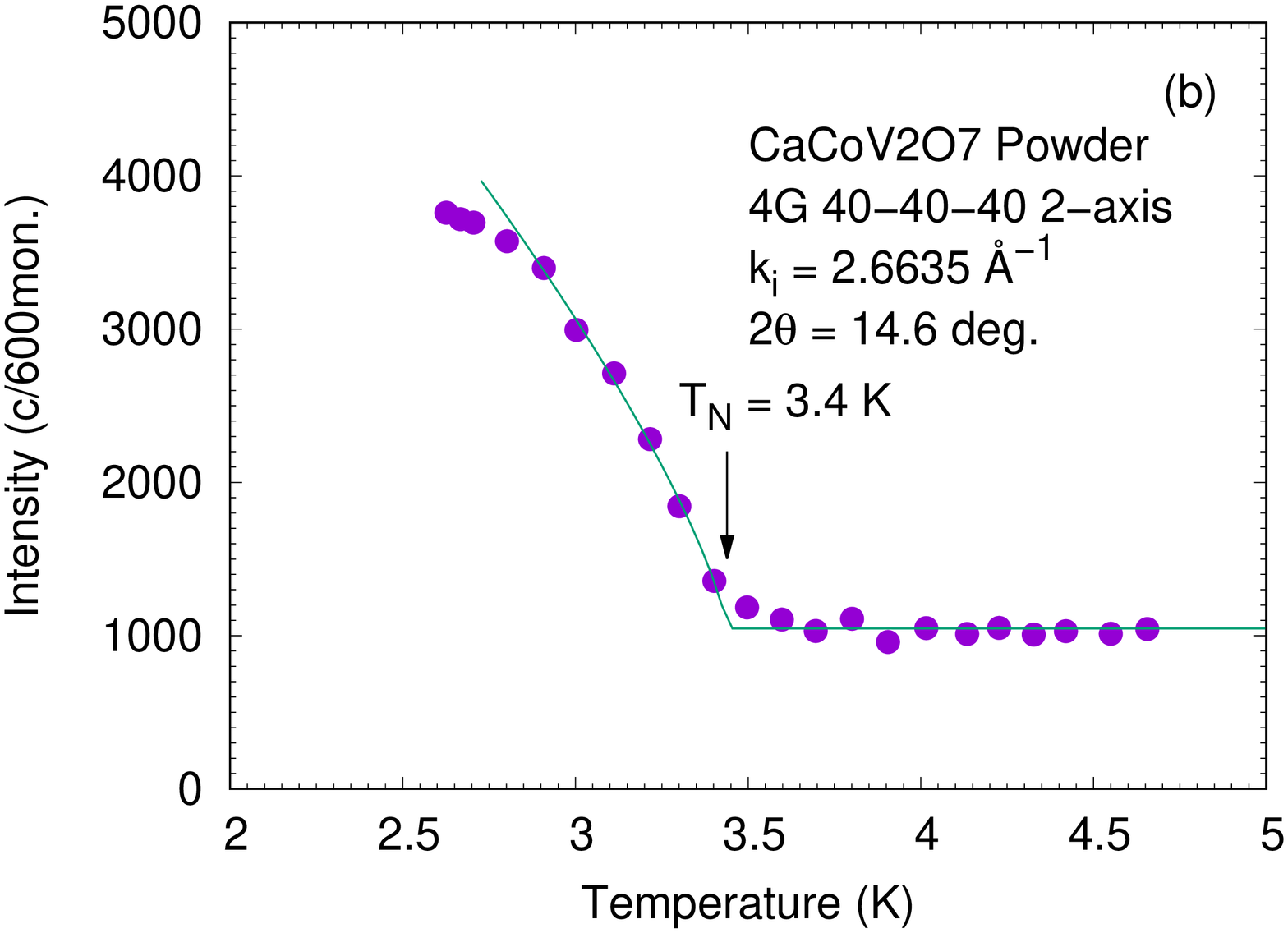}
  \includegraphics[scale=0.3, angle=-90]{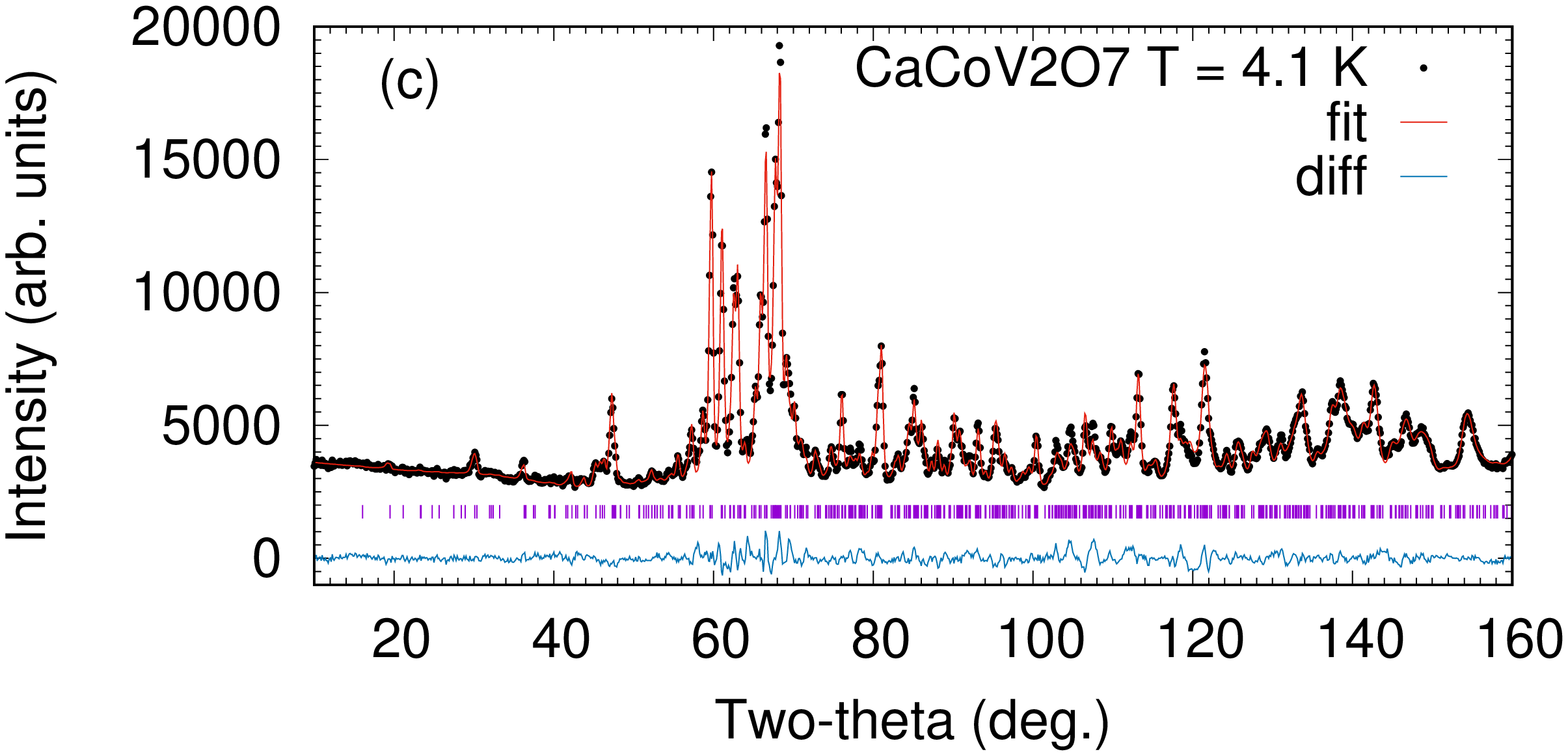}
  \includegraphics[scale=0.3, angle=-90]{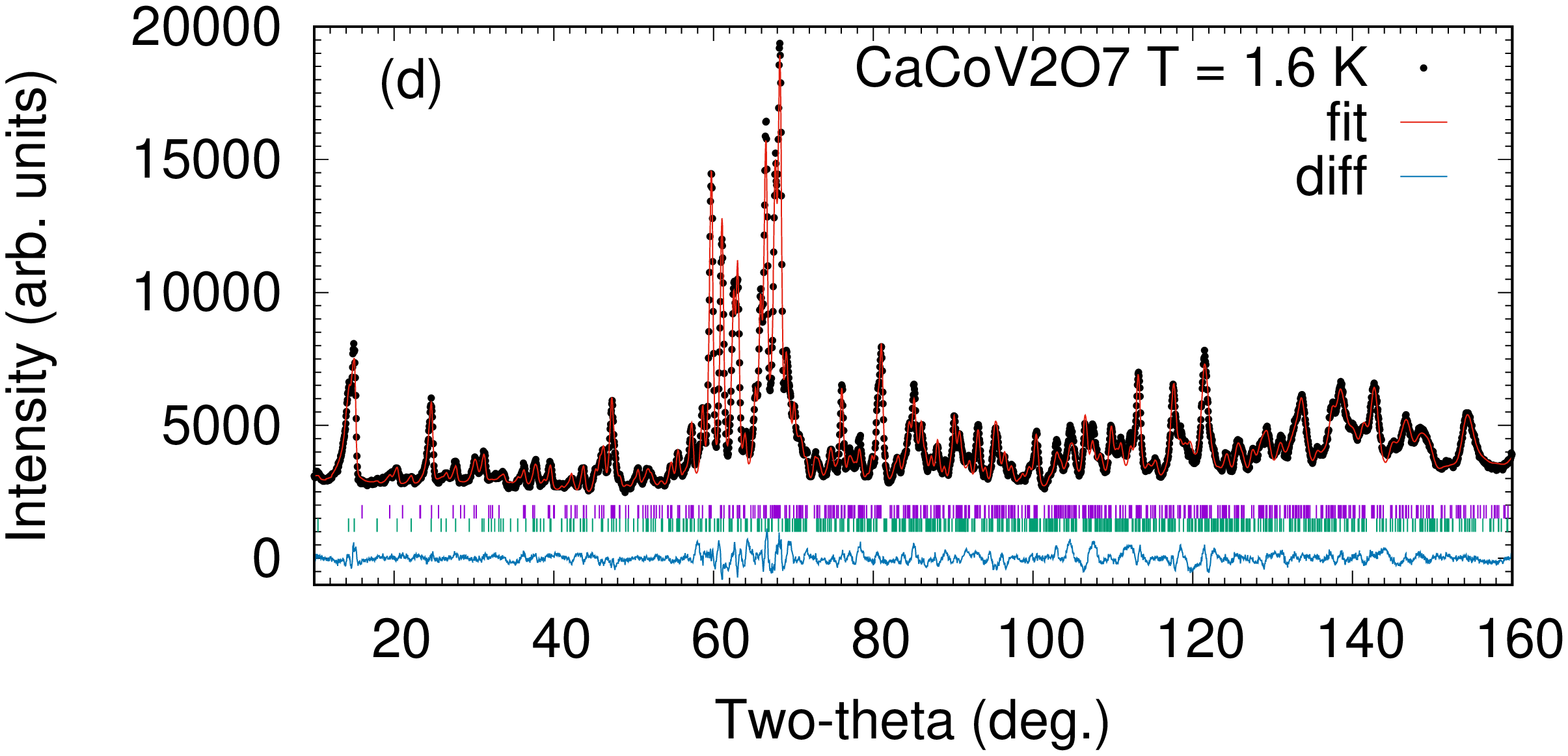}  
  \caption{\label{figure3} (a) Neutron powder diffraction patterns of CaCoV$_2$O$_7$ measured at various temperatures between 1.6~K (ordered) and 4.5~K (paramagnetic).
    (b) Temperature dependence of the integrated intensity of the lowest-$2\theta$ magnetic reflection appearing at $2\theta \simeq 15^{\circ}$.
    (c) Rietveld refinement results for the paramagnetic phase at $T = 4.1$~K.
    The vertical bars stands for the calculated nuclear Bragg reflection positions.
    (d) Rietveld refinement results for the ordered phase at $T = 1.6$~K.
    The vertical bars stands for the calculated nuclear (top) and magnetic (bottom) Bragg reflection positions.
    In (c) and (d), the difference between the observation and the calculated profile is also shown at the bottom of the figure.
  }
\end{figure*}

Neutron powder diffraction measurement was performed at various temperatures between $T = 4.5$~K and 1.6~K.
The resulting diffraction patterns  are shown in Fig.~\ref{figure3}(a).
Clearly, new peaks appear in the low-$2\theta$ region at temperature below 3.5~K, indicating that they are the magnetic Bragg peaks attributed to the long-range magnetic order detected in the macroscopic susceptibility measurement.
The temperature dependence of the lowest-$2\theta$ peak is shown in Fig.~\ref{figure3}(b).
As the temperature is lowered, the magnetic reflection intensity starts to increase; a fit to the power law $I(T) \propto (T_{\rm N} - T)^{2\beta}$ results in the estimated transition temperature of $T_{\rm N} = 3.44(2)$~K and $\beta = 0.38(5)$.
This transition temperature is in good accordance with the anomaly temperature $T_{\rm N} \sim 3.5$~K in the bulk magnetic susceptibility measurement.
The critical exponent $\beta$ is also consistent with those of 3D magnetic order~\cite{GuidaR1998}, however, the dimensionality of the order parameter could not be concluded due to insufficient temperature precision of the present experimental setup.
As per the magnetic modulation vector $\vec{q}_{\rm m}$, we found that $\vec{q}_{\rm m} = (1/2, 0, 0)$ consistently accounts for all the magnetic reflection positions.

As a representative result for the low-temperature crystal structure analysis, the neutron diffraction pattern at the paramagnetic temperature $T = 4.1$~K is analyzed using the Rietveld method starting with the reported crystallographic parameters as initial parameters.
The result of the Rietveld fitting is shown in the Fig.~\ref{figure3}(c).
The Rietveld profile with optimal parameters satisfactorily reproduces the experimental diffraction pattern, confirming that the reported crystal structure is valid at the low temperatures (but with slight change of crystallographic parameters).
Obtained optimal parameters and fractional coordinates are given in Tables~\ref{structure} and \ref{fractionalcoordinate}, respectively.
Diffraction patterns at other paramagnetic temperatures were similarly analyzed; details are given in the supplemental information, whereas representative fitting parameters are given in Table~\ref{structure}.

%\begin{table}
\begin{turnpage}
  \begin{table*}
    \caption{\label{structure} Refined crystallographic data obtained in the present powder neutron diffraction experiment.}
    \label{coeff2}
    \begin{ruledtabular}
      \begin{tabular}{lcccccccc}
        & 1.6~K & 2.1~K & 2.5~K & 2.9~K & 3.3~K & 3.7~K & 4.1~K & 4.5~K\\ 
        \hline$a$ (\AA)  & 6.7760(1)  & 6.7760(1)  & 6.7759(1)  & 6.7763(1)  & 6.7764(1)  & 6.7768(1)  & 6.7767(1)  & 6.7765(1) \\ 
        $b$ (\AA)  & 14.4450(2)  & 14.4447(4)  & 14.4451(4)  & 14.4453(4)  & 14.4463(3)  & 14.4477(4)  & 14.4472(4)  & 14.4468(3) \\ 
        $c$ (\AA)  & 11.1985(2)  & 11.1985(3)  & 11.1984(3)  & 11.1991(3)  & 11.1993(2)  & 11.2010(3)  & 11.1997(3)  & 11.1994(2) \\ 
        $\beta$ ($^\circ$)  & 100.1008(12)  & 100.1006(20)  & 100.1040(19)  & 100.1049(19)  & 100.1047(12)  & 100.1131(19)  & 100.1072(19)  & 100.1086(12) \\ 
        $V$ (\AA$^3$)  & 1079.11(3)  & 1079.09(4)  & 1079.09(4)  & 1079.23(4)  & 1079.33(3)  & 1079.65(5)  & 1079.49(5)  & 1079.39(3) \\ 
        $C_{1}^2$ ($\mu_\mathrm{B}$)  & 1.50(16)  & 1.50(25)  & 1.29(29)  & 0.85(38)  & 0.65(30)  & -- & -- & --\\ 
        $C_{2}^2$ ($\mu_\mathrm{B}$)  & $-$1.57(9)  & $-$1.80(13)  & $-$1.49(15)  & $-$0.72(18)  & $-$0.94(13)  & -- & -- & --\\ 
        $C_{3}^2$ ($\mu_\mathrm{B}$)  & 2.70(10)  & 2.76(15)  & 2.48(16)  & 2.04(19)  & 1.19(16)  & -- & -- & --\\ 
        $C_{4}^2$ ($\mu_\mathrm{B}$)  & $-$0.98(12)  & $-$0.70(18)  & $-$0.88(21)  & $-$1.40(28)  & $-$0.90(18)  & -- & -- & --\\ 
        $C_{5}^2$ ($\mu_\mathrm{B}$)  & 0.87(11)  & 0.55(18)  & 0.75(18)  & 0.96(19)  & 0.01(18)  & -- & -- & --\\ 
        $C_{6}^2$ ($\mu_\mathrm{B}$)  & $-$3.52(7)  & $-$3.33(11)  & $-$3.20(11)  & $-$2.83(14)  & $-$0.24(36)  & -- & -- & --\\ 
        Density (g/cm$^3$)  & 3.85 & 3.85 & 3.85 & 3.85 & 3.85 & 3.85 & 3.85 & 3.85\\ 
        No. of parameters\footnotemark[1]  & 84 & 84 & 84 & 84 & 84 & 78 & 78 & 78\\ 
        2$\theta$ range used for refinement ($^\circ$)  & 8.0--163.9 & 8.0--163.9 & 8.0--163.9 & 8.0--163.9 & 8.0--163.9 & 8.0--163.9 & 8.0--163.9 & 8.0--163.9\\ 
        No. of nuclear reflections & 623 & 623 & 623 & 623 & 623 & 623 & 623 & 623\\ 
        No. of magnetic reflections & 1229 & 1229 & 1229 & 1229 & 1229 & -- & -- & --\\  
        $R_\mathrm{p}$ (\%)   & 8.92 & 8.86 & 9.04 & 8.99 & 9.58 & 9.80 & 9.53 & 9.60\\ 
        $R_\mathrm{wp}$ (\%)  & 10.40 & 10.40 & 10.40 & 10.40 & 10.80 & 10.90 & 10.80 & 10.80\\
        $R_\mathrm{e}$ (\%)   & 4.38 & 4.29 & 4.38 & 4.44 & 4.68 & 4.65 & 4.57 & 4.64\\ 
        $\chi^2$  & 5.64 & 5.87 & 5.64 & 5.46 & 5.36 & 5.54 & 5.56 & 5.44\\ 
        $R_\mathrm{mag}$ (\%)  & 7.37 & 7.06 & 7.66 & 7.74 & 22.6 & -- & -- & --\\ 
      \end{tabular}
      \footnotetext[1]{Shift parameters are not included since they are not refined in the final refinement.}
    \end{ruledtabular}
  \end{table*}
\end{turnpage}

\begin{table}[t]
\caption{\label{fractionalcoordinate}Structure parameters of CaCoV$_2$O$_7$ at 4.1~K obtained in the present powder diffraction experiment.
The space group is $P2_1/c$, and the lattice parameters are $a$ = 6.7767(1) ~\AA, $b$ = 14.4472(4) ~\AA, $c$ = 11.1997(3)~\AA, and $\beta$ = 100.1072(19)~\AA.
The fractional atomic coordinates are presented.
The equivalent isotropic displacement parameters $B_\mathrm{iso}$ are listed in a unit of \AA $^2$.
Occupancy is fixed to 1 for all the atoms.}
\label{atom}
\begin{center}
\begin{tabular}{lccccc}
\hline\hline
atom & site & $x$ & $y$ & $z$ & $B_\mathrm{iso}$ \\ \hline 
Ca1 & 4e & 0.6613(25) & 0.1319(14) & 0.0177(17) & 0.86(36) \\ 
Ca2 & 4e & 0.9345(20) & 0.0017(13) & 0.7516(16) & 0.86(36) \\ 
Co1 & 4e & $-$0.0160(54) & 0.2510(23) & 0.2730(29) & 0.25(0) \\ 
Co2 & 4e & 0.3438(49) & 0.1747(22) & 0.4240(26) & 0.25(0) \\ 
V1 & 4e & 0.4906(0) & 0.2055(0) & 0.7046(0) & 0.25(0) \\ 
V2 & 4e & 0.4215(0) & 0.0139(0) & 0.6872(0) & 0.25(0) \\ 
V3 & 4e & 0.1321(0) & $-$0.1348(0) & 0.5208(0) & 0.25(0) \\ 
V4 & 4e & 0.1582(0) & 0.3723(0) & 0.5144(0) & 0.25(0) \\ 
O1 & 4e & 0.5701(18) & 0.1055(10) & 0.7923(11) & 0.33(12) \\ 
O2 & 4e & 0.3095(21) & 0.2514(10) & 0.7647(12) & 0.33(12) \\ 
O3 & 4e & 0.3564(24) & 0.2985(11) & 0.5276(12) & 0.33(12) \\ 
O4 & 4e & 0.6965(22) & 0.2714(9) & 0.7027(13) & 0.33(12) \\ 
O5 & 4e & 0.4155(21) & 0.1221(10) & 0.5845(11) & 0.33(12) \\ 
O6 & 4e & 0.6205(20) & $-$0.0520(11) & 0.6769(12) & 0.33(12) \\ 
O7 & 4e & 0.2221(19) & $-$0.0285(12) & 0.5608(12) & 0.33(12) \\ 
O8 & 4e & 0.2984(18) & $-$0.0236(9) & 0.8006(12) & 0.33(12) \\ 
O9 & 4e & $-$0.0403(23) & $-$0.1562(9) & 0.6173(14) & 0.33(12) \\ 
O10 & 4e & 0.0192(22) & $-$0.1395(10) & 0.3771(14) & 0.33(12) \\ 
O11 & 4e & 0.3244(20) & $-$0.2146(10) & 0.5509(13) & 0.33(12) \\ 
O12 & 4e & 0.0165(23) & 0.3664(9) & 0.3721(13) & 0.33(12) \\ 
O13 & 4e & $-$0.0069(24) & 0.3547(9) & 0.6223(14) & 0.33(12) \\ 
O14 & 4e & 0.2596(16) & 0.4748(11) & 0.5244(11) & 0.33(12) \\ 
\hline\hline
\end{tabular}
\end{center}
\end{table}

\begin{table}
  \caption{\label{BV1o2} List of basis vectors of the irreducible magnetic representations for the $4e$ site of $P2_1/c$ with the magnetic modulation vector $\vec{q}_{\rm m} = (1/2, 0, 0)$.
    The site index $d$ is defined as; $d = 1: (x, y, z)$, $d = 2: (-x, y+1/2, -z+1/2) + (1,0,0)$, $d = 3: (-x, -y, -z) + (1,1,1)$, and $d = 4: (x, -y + 1/2, z+1/2)$.
    The fractional coordinates $x, y$ and $z$ are the ones given in Table~\ref{atom} for the Co2 site, whereas for the Co1 site, $x$ was replaced by $x + 1$ so as the $d = 1$ site to be in the 0th unit cell.}
  \begin{ruledtabular}
    \begin{tabular}{l l l l l l l}
      IR$\nu$:$\lambda$ & $d = 1$ & $d = 2$ & $d = 3$ & $d = 4$& dim \\ \hline
      IR1:1 &  1  0  0 &	 1  0  0 &	-1  0  0 &	-1  0  0 & 1 \\
      IR1:2 & 0  1  0 &	 0 -1  0 &	 0 -1  0 &	 0  1  0 & 1 \\
      IR1:3 & 0  0  1 &	 0  0  1 &	 0  0 -1 &	 0  0 -1 & 1 \\
      IR2:1 & 1  0  0 &	-1  0  0 &	 1  0  0 &	-1  0  0 & 1\\
      IR2:2 & 0  1  0 &	 0  1  0 &	 0  1  0 &	 0  1  0 & 1\\
      IR2:3 & 0  0  1 &	 0  0 -1 &      0  0  1 &	 0  0 -1 & 1\\
      IR3:1 & 1  0  0 &	 1  0  0 &	 1  0  0 &	 1  0  0 & 1 \\
      IR3:2 & 0  1  0 &	 0 -1  0 &	 0  1  0 &	 0 -1  0 & 1 \\
      IR3:3 & 0  0  1 &	 0  0  1 &	 0  0  1 &	 0  0  1 & 1 \\
      IR4:1 & 1  0  0 &	-1  0  0 &	-1  0  0 &	 1  0  0 & 1 \\
      IR4:2 & 0  1  0 &	 0  1  0 &	 0 -1  0 &	 0 -1  0 & 1 \\
      IR4:3 & 0  0  1 &	 0  0 -1 &	 0  0 -1 &	 0  0  1 & 1 \\
    \end{tabular}
  \end{ruledtabular}
\end{table}

\begin{table}
  \caption{\label{BV0} List of basis vectors of the irreducible magnetic representations for the $4e$ site of $P2_1/c$ with the magnetic modulation vector $\vec{q}_{\rm m} = (0, 0, 0)$.
    The site index $d$ is defined as; $d = 1: (x, y, z)$, $d = 2: (-x, y+1/2, -z+1/2) + (1,0,0)$, $d = 3: (-x, -y, -z) + (1,1,1)$, and $d = 4: (x, -y + 1/2, z+1/2)$.
    The fractional coordinates $x, y$ and $z$ are the ones given in Table~\ref{atom} for the Co2 site, whereas for the Co1 site, $x$ was replaced by $x + 1$ so as the $d = 1$ site to be in the 0th unit cell.}
  \begin{ruledtabular}
    \begin{tabular}{l l l l l l l}
      IR$\nu$:$\lambda$ & $d = 1$ & $d = 2$ & $d = 3$ & $d = 4$& dim \\ \hline
      IR1:1 & 1 0 0 & -1 0 0 & 1 0 0 & -1 0 0 & 1\\
      IR1:2 & 0 1 0 & 0 1 0 & 0 1 0 & 0 1 0 & 1\\
      IR1:3 & 0 0 1 & 0 0 -1 & 0 0 1 & 0 0 -1 & 1\\
      IR2:1 &  1  0  0 &	 1  0  0 &	-1  0  0 &	-1  0  0 & 1\\
      IR2:2 &  0  1  0 &	 0 -1  0 &	 0 -1  0 &	 0  1  0& 1\\
      IR2:3 &  0  0  1 &	 0  0  1 &	 0  0 -1 &	 0  0 -1& 1\\
      IR3:1 &  1  0  0 &	-1  0  0 &	-1  0  0 &	 1  0  0& 1\\
      IR3:2 &  0  1  0 &	 0  1  0 &	 0 -1  0 &	 0 -1  0& 1\\
      IR3:3 &  0  0  1 &	 0  0 -1 &	 0  0 -1 &	 0  0  1& 1\\
      IR4:1 &  1  0  0 &	 1  0  0 &	 1  0  0 &	 1  0  0& 1\\
      IR4:2 &  0  1  0 &	 0 -1  0 &	 0  1  0 &	 0 -1  0& 1\\
      IR4:3 &  0  0  1 &	 0  0  1 &	 0  0  1 &	 0  0  1& 1\\
    \end{tabular}
  \end{ruledtabular}
\end{table}

The magnetic structure analysis was also performed using the data obtained below $T_{\rm N}$ with the aid of the magnetic representation analysis~\cite{Izyumov1979,Izyumov1991}.
In CaCoV$_2$O$_7$, there are two crystallographically inequivalent Co sites (Co1 and Co2), both of which are at the Wyckoff position $4e$.
The four symmetry related sites are not split into orbits under the $k$-group with $\vec{q}_{\rm m} = (1/2, 0, 0)$, and thus, relation between the magnetic moment directions for the four sites will be given from the magnetic representation analysis.
On the other hand, the relation between the Co1 and Co2 sites cannot be assumed from symmetry.
The basis vectors (BVs) at the $4e$ sites for each irreducible (magnetic) representation (IR) of the $k$-group with $\vec{q}_{\rm m} = (1/2, 0, 0)$ are given in Table~\ref{BV1o2}.
The magnetic representation includes four one-dimensional IRs (IR1 to IR4), each of which has three BVs.
It should be noted that all the IRs listed in Table~\ref{BV1o2} correspond to four magnetic space groups $P_a2_1/c$ (14.80; BNS~\cite{BelovNV1957}) with different relationships to the parent $P2_1/c$ space group.

\begin{figure}
  \includegraphics[scale=0.35, angle=90, trim=0 0 0 100]{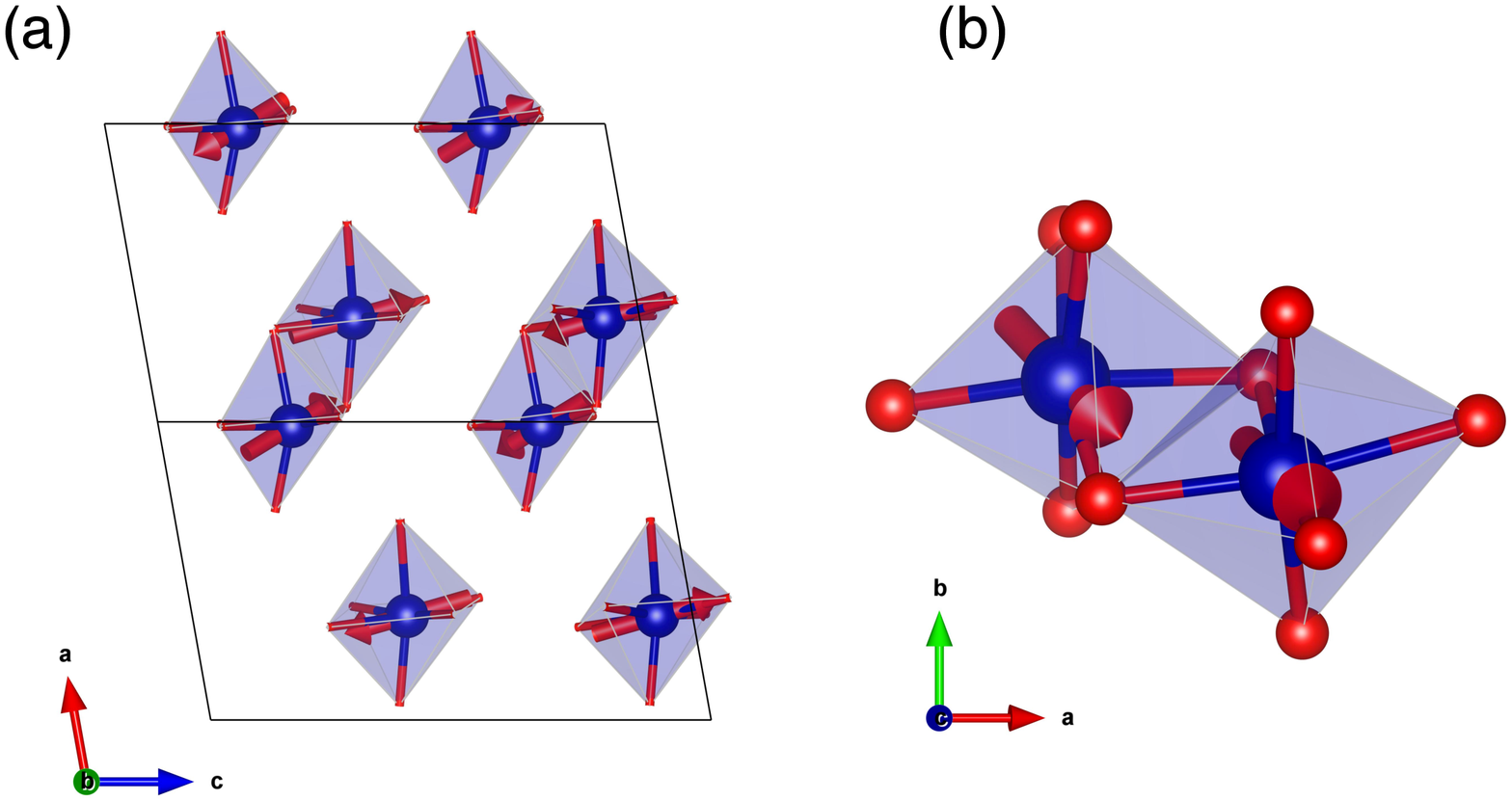}
  \caption{\label{figure4} (a) Presently proposed magnetic structure for CaCoV$_2$O$_7$ at the base temperature.
    Magnetic moments in the half of the unit cell along the $b$-axis is drawn.
    (b) Ordered moment directions of the Co$^{2+}$ ions in the edge-sharing Co-octahedra.
    Magnetic moment heads to the closest vertex oxygen.
    Structure illustrations were made with VESTA~\cite{MommaK2011}.    
  }
\end{figure}

The magnetic structure is assumed to be represented by the single magnetic modulation $\vec{q}_{\rm m}$ as:
\begin{equation}
  \vec{S}_{\vec{l}, \vec{d}} = \frac{S}{2} \left \{
    \vec{a}_{\vec{d}} \exp \left [- {\rm i} \vec{q}_{\rm m} 
      \cdot \vec{l} \right] 
    + {\rm c.c.}\right \},
\end{equation}
where $\vec{S}_{\vec{l}, \vec{d}}$ stands for the spin direction (vector) at the $d$-th site (at the site fractional coordinate $\vec{d}$) in the $l$-th unit cell (at the position $\vec{l})$.
Following the Landau theory, we assumed that one of the four IRs will be selected through the continuous phase transition.
Accordingly, the Fourier component of the magnetic moment vectors at the $d$-th site, $\vec{a}_{\vec{d}}$, is assumed to be a linear combination of the three BVs $\vec{\Psi}^{\vec{d}}_{\nu, \lambda}$ of a single ($\nu$-th) IR for each Co$^{2+}$ sites as:
\begin{equation}
  \vec{a}_{\vec{d}} = \sum_{\lambda} C_{\lambda}^{\nu} \vec{\Psi}^{\vec{d}}_{\nu, \lambda},
\end{equation}
where $C_{\lambda}^{\nu}$ is the linear combination coefficient for a given BV, and $\lambda$ numbers each BV.
In total six $C_{\lambda}^{\nu}$ parameters are used in the fitting; first three ($\lambda = 1,...,3$) are for the BV1 to BV3 at the Co1 site, whereas the last three ($\lambda = 4,...,6$) for those at Co2.
Crystal structure parameters are also refined at each temperature.
We have tried all the IRs in the fitting procedure, and found that BVs belonging to IR2 reproduce the observed magnetic reflection intensity satisfactorily.
We note that the IR2 corresponds to $mY2-$ magnetic IR in the Miller and Love notation~\cite{MillerSC1967}.
The optimal $C_{\lambda}^{\nu = 2}$ coefficients at each temperature are given in Table~\ref{structure} together with fitting conditions and representative crystallographic parameters.
The magnetic structure at the base temperature $T = 1.6$~K is depicted in Fig.~\ref{figure4}.

From the $C$ coefficients, we estimate the sizes of the ordered moments for the Co1 and Co2 sites as $m_{\rm Co1} \simeq 3.4~\mu_{\rm B}$ and $m_{\rm Co2} \simeq 3.1~\mu_{\rm B}$ at $T = 1.6$~K, respectively.
These ordered moment sizes are significantly reduced from the effective moment size estimated from the high-temperature Curie-Weiss fit.
Such a reduction of the magnetic moment size at low temperatures was not unusual in the Co$^{2+}$ ions in the nearly octahedral coordinate, as exemplified by the ordered double perovskite Sr$_2$CoTeO$_6$~\cite{MartinOS2005}; in this compound, at high temperatures the effective moment is estimated as 5.46~$\mu_{\rm B}$, whereas the ordered magnetic moment estimated at low temperatures is 2.25~$\mu_{\rm B}$, being consistent with the theoretical moment size of the ground doublet due to the spin-orbit coupling.
It should be noted, however, that in the present CaCoV$_2$O$_7$, the low-temperature ordered moment size is larger than the one observed in Sr$_2$CoTeO$_6$.
We speculate that this may result from the significantly distorted octahedral coordinate in the present CaCoV$_2$O$_7$, which may mix the higher energy CEF states into the ground doublet.

The obtained magnetic structure gives us several clues to understand the magnetism of the double pyrovanadate CaCoV$_2$O$_7$.
First, the two magnetic moments in the edge-sharing octahedra are almost parallel to each other, forming a ferromagnetic spin dimer.
Having in mind that they are inequivalent sites, for which spin directions are not fixed by the symmetry, this result indicates that the intra-dimer ferromagnetic coupling is in effect, and bulk antiferromagnetic order is indeed due to inter-dimer antiferromagnetic interactions.
It should be noted that a formation of ferromagnetic dimer for the Co$^{2+}$ ions is rather special; rare examples may be found in dinuclear Co$^{2+}$ complexes~\cite{LarrabeeJA2008,DaumannLJ2013}.
We think the ferromagnetic interactions between the Co1 and Co2 atoms are due to two Co1-O-Co2 super-exchange paths within the dimer; in CaCoV$_2$O$_7$ the Co1-O-Co2 bond angles are 95.21$^{\circ}$ and 99.29$^{\circ}$~\cite{MurashovaEV1993}.
These values do not contradict with the Co-O-Co bond angles observed for the ferromagnetic bonds in the Co$^{2+}$ complexes~\cite{LarrabeeJA2008,DaumannLJ2013}.

Secondly, the magnetic structure obtained from the neutron diffraction is strictly antiferromagnetic; because of the alternating moment configuration due to the half magnetic modulation $\vec{q}_{\rm m} = (1/2, 0, 0 )$, even the ferromagnetic BV (e.g. IR2:2) would be compensated by the moments in the $\vec{a}$-translated unit cell.
This is apparently inconsistent with the appearance of the weak ferromagnetic component detected in the bulk magnetic measurements.
We note that the translational coset representatives for the $k$-group of $P2_1/c$ with $\vec{k} = \vec{q}_{\rm m}$ are identical to those of the (translational coset) representatives of the nonmagnetic space group $P2_1/c$ (with $\vec{k} = (0,0,0)$).
The observed magnetic structure (IR2) is invariant under the symmetry operations 1 and $c$.
(The other two operations ($2(0,1/2,0)$ and $\bar{1}$) flip the moment direction.)
This infers that any other IRs that are invariant under 1 and $c$ may couple to the observed magnetic structure as a secondary order parameter.
For $\vec{k} = (0,0,0)$ with which the bulk magnetic moment may be allowed, we found that IR1 and IR2 of $P2_1/c$ satisfy this condition.
Among them, IR1 is compatible with the bulk ferromagnetic moment along the $b$-axis, whereas IR2 are strictly antiferromagnetic.
(See Table~\ref{BV0} for the BVs of IRs for $\vec{k} = (0,0,0)$.)
Therefore, we speculate that the weak-ferromagnetic moment along the $b$-axis accompanies the dominant antiferromagnetic order in CaCoV$_2$O$_7$.
It is noted here that judging from the size of the ferromagnetic component ($\sim 0.05~\mu_{\rm B}$), it is certainly impossible to detect the ferromagnetic moment using the powder diffraction, where the magnetic component appears on top of the much stronger nuclear reflections.
On the other hand, the moment direction may be concluded using the single crystal magnetization measurement.
Such experiment is, therefore, desired in future to confirm the above speculation from the symmetry analysis.

Another possible mechanism for the weak ferromagnetic moment may be impurity or surface/interface effect.
It is reported that in the nanoparticle Co$_3$O$_4$, a weak net ferromagnetic moment was observed due to the interface effect between the surface ferromagnetic-like spins covering the core antiferromagnetically ordered spins~\cite{DuttaP2007}.
This Co$_3$O$_4$ phase may be included in the present sample as an impurity secondary phase; it is noted that such an impurity secondary phase is certainly below the detection limit of the present X-ray or neutron powder diffraction judging from the net ferromagnetic moment size (0.05~$\mu_{\rm B}$).
Much intriguing possibility may be that the surface/interface effect proposed in Co$_3$O$_4$ indeed works in the present polycrystalline CaCoV$_2$O$_7$; for such a case it is naturally explained that the weak ferromagnetic moment simultaneously appears at the antiferromagnetic transition temperature.
Again, the single crystal magnetic measurements will clarify this possibility, which is left for the future study.

Finally, we note that the oxygen tetrahedra around the Co1 and Co2 sites are highly distorted upon magnetic ordering.
The Co-O distances distribute between 1.845~\AA\ to 2.302~\AA\ at the lowest temperature, being a significantly larger distribution than the room temperature one (1.975~\AA\ to 2.225~\AA.)
Furthermore, a close relation between the distortion of the octahedra and magnetic moment direction can be found; the magnetic moment points toward the vertex oxygen ion that is closest to the Co$^{2+}$ ion.
This indicates that strong CEF is in effect to fix the magnetic moment direction.

\section{Summary}
Bulk magnetization, X-ray, and neutron diffraction measurements have been performed on the double pyrovanadate CaCoV$_2$O$_7$ using the powder sample.
The crystal structure at the room temperature, as well as the base temperature, is confirmed to be monoclinic $P2_1/c$, as reported earlier.
The inverse magnetic susceptibility in the wide temperature range $100< T < 300$~K exhibits the linear Curie-Weiss behavior.
The Weiss temperature was estimated to be $\simeq -22.5$~K, indicating dominant antiferromagnetic interactions, whereas the estimated effective magnetic moment is $\simeq 5.5~\mu_{\rm B}$, indicating considerable orbital contribution for the present Co$^{2+}$ ions.
With further lowering temperature, the long-range antiferromagnetic order, accompanied by the weak ferromagnetic component of $\sim 0.05~\mu_{\rm B}$, takes place.
Neutron diffraction reveals that the magnetic transition temperature is $T_{\rm N} = 3.44$~K and the magnetic modulation vector is $\vec{q}_{\rm m} = (1/2, 0, 0)$.
From the magnetic structure analysis, we concluded that the two nearest-neighbor Co1 and Co2 moments in the edge-sharing adjacent octahedra are almost parallel, indicating the formation of ferromagnetic dimers.
The ferromagnetic dimers in turn form antiferromagnetic long-range order due to the sizable inter-dimer antiferromagnetic interactions.
From the symmetry analysis, it is speculated that the antiferromagnetic order may be accompanied by the weak ferromagnetic moment along the $b$-axis, via weak coupling to IR corresponding to $\vec{k} = (0, 0, 0)$.

{\it Note added:} During the final preparation of the present manuscript, we became aware of one presentation at the annual meeting of the Physical Society of Japan on the magnetism of the title compound CaCoV$_2$O$_7$~\cite{YogiAK2021}.
While focus of the above work was on the magnetic phase diagram, which is different from ours, magnetism at low magnetic fields and magnetic structures were consistent with the results reported in the present manuscript.

\begin{acknowledgments}
  This work is partly supported by grants-in-aids for scientific research (JP19H01834, JP19K21839, JP19H05824, 19KK0069) from MEXT of Japan, and by the research program ``dynamic alliance for open innovation bridging human, environment, and materials.''
  The experiments at JRR-3, as well as the travel expenses for the neutron scattering experiment at ANSTO, was partly supported by the General User Program for Neutron Scattering Experiments, Institute for Solid State Physics, University of Tokyo.
\end{acknowledgments}

\bibliography{cacov2o7}% Produces the bibliography via BibTeX.

%merlin.mbs apsrev4-1.bst 2010-07-25 4.21a (PWD, AO, DPC) hacked
%Control: key (0)
%Control: author (72) initials jnrlst
%Control: editor formatted (1) identically to author
%Control: production of article title (-1) disabled
%Control: page (0) single
%Control: year (1) truncated
%Control: production of eprint (0) enabled
\begin{thebibliography}{51}%
\makeatletter
\providecommand \@ifxundefined [1]{%
 \@ifx{#1\undefined}
}%
\providecommand \@ifnum [1]{%
 \ifnum #1\expandafter \@firstoftwo
 \else \expandafter \@secondoftwo
 \fi
}%
\providecommand \@ifx [1]{%
 \ifx #1\expandafter \@firstoftwo
 \else \expandafter \@secondoftwo
 \fi
}%
\providecommand \natexlab [1]{#1}%
\providecommand \enquote  [1]{``#1''}%
\providecommand \bibnamefont  [1]{#1}%
\providecommand \bibfnamefont [1]{#1}%
\providecommand \citenamefont [1]{#1}%
\providecommand \href@noop [0]{\@secondoftwo}%
\providecommand \href [0]{\begingroup \@sanitize@url \@href}%
\providecommand \@href[1]{\@@startlink{#1}\@@href}%
\providecommand \@@href[1]{\endgroup#1\@@endlink}%
\providecommand \@sanitize@url [0]{\catcode `\\12\catcode `\$12\catcode
  `\&12\catcode `\#12\catcode `\^12\catcode `\_12\catcode `\%12\relax}%
\providecommand \@@startlink[1]{}%
\providecommand \@@endlink[0]{}%
\providecommand \url  [0]{\begingroup\@sanitize@url \@url }%
\providecommand \@url [1]{\endgroup\@href {#1}{\urlprefix }}%
\providecommand \urlprefix  [0]{URL }%
\providecommand \Eprint [0]{\href }%
\providecommand \doibase [0]{http://dx.doi.org/}%
\providecommand \selectlanguage [0]{\@gobble}%
\providecommand \bibinfo  [0]{\@secondoftwo}%
\providecommand \bibfield  [0]{\@secondoftwo}%
\providecommand \translation [1]{[#1]}%
\providecommand \BibitemOpen [0]{}%
\providecommand \bibitemStop [0]{}%
\providecommand \bibitemNoStop [0]{.\EOS\space}%
\providecommand \EOS [0]{\spacefactor3000\relax}%
\providecommand \BibitemShut  [1]{\csname bibitem#1\endcsname}%
\let\auto@bib@innerbib\@empty
%</preamble>
\bibitem [{\citenamefont {Mercurio-Lavaud}\ and\ \citenamefont
  {Frit}(1973)}]{MercurioLavaud1973}%
  \BibitemOpen
  \bibfield  {author} {\bibinfo {author} {\bibfnamefont {D.}~\bibnamefont
  {Mercurio-Lavaud}}\ and\ \bibinfo {author} {\bibfnamefont {B.}~\bibnamefont
  {Frit}},\ }\href {\doibase 10.1107/S0567740873007478} {\bibfield  {journal}
  {\bibinfo  {journal} {Acta Crystallographica Section B}\ }\textbf {\bibinfo
  {volume} {29}},\ \bibinfo {pages} {2737} (\bibinfo {year}
  {1973})}\BibitemShut {NoStop}%
\bibitem [{\citenamefont {Calvo}\ and\ \citenamefont
  {Faggiani}(1975{\natexlab{a}})}]{Calvo1975}%
  \BibitemOpen
  \bibfield  {author} {\bibinfo {author} {\bibfnamefont {C.}~\bibnamefont
  {Calvo}}\ and\ \bibinfo {author} {\bibfnamefont {R.}~\bibnamefont
  {Faggiani}},\ }\href {\doibase 10.1107/S0567740875003354} {\bibfield
  {journal} {\bibinfo  {journal} {Acta Crystallographica Section B}\ }\textbf
  {\bibinfo {volume} {31}},\ \bibinfo {pages} {603} (\bibinfo {year}
  {1975}{\natexlab{a}})}\BibitemShut {NoStop}%
\bibitem [{\citenamefont {Liao}\ \emph {et~al.}(1996)\citenamefont {Liao},
  \citenamefont {Leroux}, \citenamefont {Piffard}, \citenamefont {Guyomard},\
  and\ \citenamefont {Payen}}]{Liao1996}%
  \BibitemOpen
  \bibfield  {author} {\bibinfo {author} {\bibfnamefont {J.-H.}\ \bibnamefont
  {Liao}}, \bibinfo {author} {\bibfnamefont {F.}~\bibnamefont {Leroux}},
  \bibinfo {author} {\bibfnamefont {Y.}~\bibnamefont {Piffard}}, \bibinfo
  {author} {\bibfnamefont {D.}~\bibnamefont {Guyomard}}, \ and\ \bibinfo
  {author} {\bibfnamefont {C.}~\bibnamefont {Payen}},\ }\href {\doibase
  https://doi.org/10.1006/jssc.1996.0030} {\bibfield  {journal} {\bibinfo
  {journal} {Journal of Solid State Chemistry}\ }\textbf {\bibinfo {volume}
  {121}},\ \bibinfo {pages} {214} (\bibinfo {year} {1996})}\BibitemShut
  {NoStop}%
\bibitem [{\citenamefont {Touaiher}\ \emph {et~al.}(2004)\citenamefont
  {Touaiher}, \citenamefont {Rissouli}, \citenamefont {Benkhouja},
  \citenamefont {Taibi}, \citenamefont {Aride}, \citenamefont {Boukhari},\ and\
  \citenamefont {Heulin}}]{Touaiher2004}%
  \BibitemOpen
  \bibfield  {author} {\bibinfo {author} {\bibfnamefont {M.}~\bibnamefont
  {Touaiher}}, \bibinfo {author} {\bibfnamefont {K.}~\bibnamefont {Rissouli}},
  \bibinfo {author} {\bibfnamefont {K.}~\bibnamefont {Benkhouja}}, \bibinfo
  {author} {\bibfnamefont {M.}~\bibnamefont {Taibi}}, \bibinfo {author}
  {\bibfnamefont {J.}~\bibnamefont {Aride}}, \bibinfo {author} {\bibfnamefont
  {A.}~\bibnamefont {Boukhari}}, \ and\ \bibinfo {author} {\bibfnamefont
  {B.}~\bibnamefont {Heulin}},\ }\href {\doibase
  https://doi.org/10.1016/j.matchemphys.2003.11.032} {\bibfield  {journal}
  {\bibinfo  {journal} {Materials Chemistry and Physics}\ }\textbf {\bibinfo
  {volume} {85}},\ \bibinfo {pages} {41} (\bibinfo {year} {2004})}\BibitemShut
  {NoStop}%
\bibitem [{\citenamefont {Krivovichev}\ \emph {et~al.}(2005)\citenamefont
  {Krivovichev}, \citenamefont {Filatov}, \citenamefont {Cherepansky},
  \citenamefont {Armbruster},\ and\ \citenamefont
  {Pankratova}}]{KrivovichevSV2005}%
  \BibitemOpen
  \bibfield  {author} {\bibinfo {author} {\bibfnamefont {S.~V.}\ \bibnamefont
  {Krivovichev}}, \bibinfo {author} {\bibfnamefont {S.~K.}\ \bibnamefont
  {Filatov}}, \bibinfo {author} {\bibfnamefont {P.~N.}\ \bibnamefont
  {Cherepansky}}, \bibinfo {author} {\bibfnamefont {T.}~\bibnamefont
  {Armbruster}}, \ and\ \bibinfo {author} {\bibfnamefont {O.~Y.}\ \bibnamefont
  {Pankratova}},\ }\href@noop {} {\bibfield  {journal} {\bibinfo  {journal}
  {Can. Mineral.}\ }\textbf {\bibinfo {volume} {43}},\ \bibinfo {pages} {671}
  (\bibinfo {year} {2005})}\BibitemShut {NoStop}%
\bibitem [{\citenamefont {He}\ and\ \citenamefont {Ueda}(2008)}]{HeZ2008}%
  \BibitemOpen
  \bibfield  {author} {\bibinfo {author} {\bibfnamefont {Z.}~\bibnamefont
  {He}}\ and\ \bibinfo {author} {\bibfnamefont {Y.}~\bibnamefont {Ueda}},\
  }\href {\doibase https://doi.org/10.1016/j.jssc.2007.11.028} {\bibfield
  {journal} {\bibinfo  {journal} {Journal of Solid State Chemistry}\ }\textbf
  {\bibinfo {volume} {181}},\ \bibinfo {pages} {235} (\bibinfo {year}
  {2008})}\BibitemShut {NoStop}%
\bibitem [{\citenamefont {Tsirlin}\ \emph {et~al.}(2010)\citenamefont
  {Tsirlin}, \citenamefont {Janson},\ and\ \citenamefont
  {Rosner}}]{TsirlinAA2010}%
  \BibitemOpen
  \bibfield  {author} {\bibinfo {author} {\bibfnamefont {A.~A.}\ \bibnamefont
  {Tsirlin}}, \bibinfo {author} {\bibfnamefont {O.}~\bibnamefont {Janson}}, \
  and\ \bibinfo {author} {\bibfnamefont {H.}~\bibnamefont {Rosner}},\ }\href
  {\doibase 10.1103/PhysRevB.82.144416} {\bibfield  {journal} {\bibinfo
  {journal} {Phys. Rev. B}\ }\textbf {\bibinfo {volume} {82}},\ \bibinfo
  {pages} {144416} (\bibinfo {year} {2010})}\BibitemShut {NoStop}%
\bibitem [{\citenamefont {Sannigrahi}\ \emph {et~al.}(2017)\citenamefont
  {Sannigrahi}, \citenamefont {Giri},\ and\ \citenamefont
  {Majumdar}}]{SannigrahiJ2017}%
  \BibitemOpen
  \bibfield  {author} {\bibinfo {author} {\bibfnamefont {J.}~\bibnamefont
  {Sannigrahi}}, \bibinfo {author} {\bibfnamefont {S.}~\bibnamefont {Giri}}, \
  and\ \bibinfo {author} {\bibfnamefont {S.}~\bibnamefont {Majumdar}},\ }\href
  {\doibase https://doi.org/10.1016/j.jpcs.2016.09.017} {\bibfield  {journal}
  {\bibinfo  {journal} {Journal of Physics and Chemistry of Solids}\ }\textbf
  {\bibinfo {volume} {101}},\ \bibinfo {pages} {1 } (\bibinfo {year}
  {2017})}\BibitemShut {NoStop}%
\bibitem [{\citenamefont {Bhowal}\ \emph {et~al.}(2017)\citenamefont {Bhowal},
  \citenamefont {Sannigrahi}, \citenamefont {Majumdar},\ and\ \citenamefont
  {Dasgupta}}]{BhowalS2017}%
  \BibitemOpen
  \bibfield  {author} {\bibinfo {author} {\bibfnamefont {S.}~\bibnamefont
  {Bhowal}}, \bibinfo {author} {\bibfnamefont {J.}~\bibnamefont {Sannigrahi}},
  \bibinfo {author} {\bibfnamefont {S.}~\bibnamefont {Majumdar}}, \ and\
  \bibinfo {author} {\bibfnamefont {I.}~\bibnamefont {Dasgupta}},\ }\href
  {\doibase 10.1103/PhysRevB.95.075110} {\bibfield  {journal} {\bibinfo
  {journal} {Phys. Rev. B}\ }\textbf {\bibinfo {volume} {95}},\ \bibinfo
  {pages} {075110} (\bibinfo {year} {2017})}\BibitemShut {NoStop}%
\bibitem [{\citenamefont {Ji}\ \emph {et~al.}(2019)\citenamefont {Ji},
  \citenamefont {Yin}, \citenamefont {Zhu}, \citenamefont {Kumar},
  \citenamefont {Li}, \citenamefont {Li}, \citenamefont {Jin}, \citenamefont
  {Nandi}, \citenamefont {Sun}, \citenamefont {Su}, \citenamefont {Br\"uckel},
  \citenamefont {Lee}, \citenamefont {Harmon}, \citenamefont {Ke},
  \citenamefont {Ouyang},\ and\ \citenamefont {Xiao}}]{JiWH2019}%
  \BibitemOpen
  \bibfield  {author} {\bibinfo {author} {\bibfnamefont {W.~H.}\ \bibnamefont
  {Ji}}, \bibinfo {author} {\bibfnamefont {L.}~\bibnamefont {Yin}}, \bibinfo
  {author} {\bibfnamefont {W.~M.}\ \bibnamefont {Zhu}}, \bibinfo {author}
  {\bibfnamefont {C.~M.~N.}\ \bibnamefont {Kumar}}, \bibinfo {author}
  {\bibfnamefont {C.}~\bibnamefont {Li}}, \bibinfo {author} {\bibfnamefont
  {H.-F.}\ \bibnamefont {Li}}, \bibinfo {author} {\bibfnamefont {W.~T.}\
  \bibnamefont {Jin}}, \bibinfo {author} {\bibfnamefont {S.}~\bibnamefont
  {Nandi}}, \bibinfo {author} {\bibfnamefont {X.}~\bibnamefont {Sun}}, \bibinfo
  {author} {\bibfnamefont {Y.}~\bibnamefont {Su}}, \bibinfo {author}
  {\bibfnamefont {T.}~\bibnamefont {Br\"uckel}}, \bibinfo {author}
  {\bibfnamefont {Y.}~\bibnamefont {Lee}}, \bibinfo {author} {\bibfnamefont
  {B.~N.}\ \bibnamefont {Harmon}}, \bibinfo {author} {\bibfnamefont
  {L.}~\bibnamefont {Ke}}, \bibinfo {author} {\bibfnamefont {Z.~W.}\
  \bibnamefont {Ouyang}}, \ and\ \bibinfo {author} {\bibfnamefont
  {Y.}~\bibnamefont {Xiao}},\ }\href {\doibase 10.1103/PhysRevB.100.134420}
  {\bibfield  {journal} {\bibinfo  {journal} {Phys. Rev. B}\ }\textbf {\bibinfo
  {volume} {100}},\ \bibinfo {pages} {134420} (\bibinfo {year}
  {2019})}\BibitemShut {NoStop}%
\bibitem [{\citenamefont {Yin}\ \emph {et~al.}(2019)\citenamefont {Yin},
  \citenamefont {Ouyang}, \citenamefont {Wang}, \citenamefont {Yue},
  \citenamefont {Chen}, \citenamefont {He}, \citenamefont {Wang}, \citenamefont
  {Xia},\ and\ \citenamefont {Liu}}]{YinL2019}%
  \BibitemOpen
  \bibfield  {author} {\bibinfo {author} {\bibfnamefont {L.}~\bibnamefont
  {Yin}}, \bibinfo {author} {\bibfnamefont {Z.~W.}\ \bibnamefont {Ouyang}},
  \bibinfo {author} {\bibfnamefont {J.~F.}\ \bibnamefont {Wang}}, \bibinfo
  {author} {\bibfnamefont {X.~Y.}\ \bibnamefont {Yue}}, \bibinfo {author}
  {\bibfnamefont {R.}~\bibnamefont {Chen}}, \bibinfo {author} {\bibfnamefont
  {Z.~Z.}\ \bibnamefont {He}}, \bibinfo {author} {\bibfnamefont {Z.~X.}\
  \bibnamefont {Wang}}, \bibinfo {author} {\bibfnamefont {Z.~C.}\ \bibnamefont
  {Xia}}, \ and\ \bibinfo {author} {\bibfnamefont {Y.}~\bibnamefont {Liu}},\
  }\href {\doibase 10.1103/PhysRevB.99.134434} {\bibfield  {journal} {\bibinfo
  {journal} {Phys. Rev. B}\ }\textbf {\bibinfo {volume} {99}},\ \bibinfo
  {pages} {134434} (\bibinfo {year} {2019})}\BibitemShut {NoStop}%
\bibitem [{\citenamefont {Sannigrahi}\ \emph {et~al.}(2015)\citenamefont
  {Sannigrahi}, \citenamefont {Bhowal}, \citenamefont {Giri}, \citenamefont
  {Majumdar},\ and\ \citenamefont {Dasgupta}}]{Sannigrahi2015}%
  \BibitemOpen
  \bibfield  {author} {\bibinfo {author} {\bibfnamefont {J.}~\bibnamefont
  {Sannigrahi}}, \bibinfo {author} {\bibfnamefont {S.}~\bibnamefont {Bhowal}},
  \bibinfo {author} {\bibfnamefont {S.}~\bibnamefont {Giri}}, \bibinfo {author}
  {\bibfnamefont {S.}~\bibnamefont {Majumdar}}, \ and\ \bibinfo {author}
  {\bibfnamefont {I.}~\bibnamefont {Dasgupta}},\ }\href {\doibase
  10.1103/PhysRevB.91.220407} {\bibfield  {journal} {\bibinfo  {journal} {Phys.
  Rev. B}\ }\textbf {\bibinfo {volume} {91}},\ \bibinfo {pages} {220407}
  (\bibinfo {year} {2015})}\BibitemShut {NoStop}%
\bibitem [{\citenamefont {Lee}\ \emph {et~al.}(2016)\citenamefont {Lee},
  \citenamefont {Jang}, \citenamefont {Dissanayake}, \citenamefont {Lee},\ and\
  \citenamefont {Jeong}}]{LeeYW2016}%
  \BibitemOpen
  \bibfield  {author} {\bibinfo {author} {\bibfnamefont {Y.-W.}\ \bibnamefont
  {Lee}}, \bibinfo {author} {\bibfnamefont {T.-H.}\ \bibnamefont {Jang}},
  \bibinfo {author} {\bibfnamefont {S.~E.}\ \bibnamefont {Dissanayake}},
  \bibinfo {author} {\bibfnamefont {S.}~\bibnamefont {Lee}}, \ and\ \bibinfo
  {author} {\bibfnamefont {Y.~H.}\ \bibnamefont {Jeong}},\ }\href {\doibase
  10.1209/0295-5075/113/27007} {\bibfield  {journal} {\bibinfo  {journal}
  {{EPL} (Europhysics Letters)}\ }\textbf {\bibinfo {volume} {113}},\ \bibinfo
  {pages} {27007} (\bibinfo {year} {2016})}\BibitemShut {NoStop}%
\bibitem [{\citenamefont {Chen}\ \emph {et~al.}(2018)\citenamefont {Chen},
  \citenamefont {Wang}, \citenamefont {Ouyang}, \citenamefont {He},
  \citenamefont {Wang}, \citenamefont {Lin}, \citenamefont {Liu}, \citenamefont
  {Lu}, \citenamefont {Liu}, \citenamefont {Dong}, \citenamefont {Liu},
  \citenamefont {Xia}, \citenamefont {Matsuo}, \citenamefont {Kohama},\ and\
  \citenamefont {Kindo}}]{ChenR2018}%
  \BibitemOpen
  \bibfield  {author} {\bibinfo {author} {\bibfnamefont {R.}~\bibnamefont
  {Chen}}, \bibinfo {author} {\bibfnamefont {J.~F.}\ \bibnamefont {Wang}},
  \bibinfo {author} {\bibfnamefont {Z.~W.}\ \bibnamefont {Ouyang}}, \bibinfo
  {author} {\bibfnamefont {Z.~Z.}\ \bibnamefont {He}}, \bibinfo {author}
  {\bibfnamefont {S.~M.}\ \bibnamefont {Wang}}, \bibinfo {author}
  {\bibfnamefont {L.}~\bibnamefont {Lin}}, \bibinfo {author} {\bibfnamefont
  {J.~M.}\ \bibnamefont {Liu}}, \bibinfo {author} {\bibfnamefont {C.~L.}\
  \bibnamefont {Lu}}, \bibinfo {author} {\bibfnamefont {Y.}~\bibnamefont
  {Liu}}, \bibinfo {author} {\bibfnamefont {C.}~\bibnamefont {Dong}}, \bibinfo
  {author} {\bibfnamefont {C.~B.}\ \bibnamefont {Liu}}, \bibinfo {author}
  {\bibfnamefont {Z.~C.}\ \bibnamefont {Xia}}, \bibinfo {author} {\bibfnamefont
  {A.}~\bibnamefont {Matsuo}}, \bibinfo {author} {\bibfnamefont
  {Y.}~\bibnamefont {Kohama}}, \ and\ \bibinfo {author} {\bibfnamefont
  {K.}~\bibnamefont {Kindo}},\ }\href {\doibase 10.1103/PhysRevB.98.184404}
  {\bibfield  {journal} {\bibinfo  {journal} {Phys. Rev. B}\ }\textbf {\bibinfo
  {volume} {98}},\ \bibinfo {pages} {184404} (\bibinfo {year}
  {2018})}\BibitemShut {NoStop}%
\bibitem [{\citenamefont {Chen}\ \emph {et~al.}(2019)\citenamefont {Chen},
  \citenamefont {Wang}, \citenamefont {Ouyang}, \citenamefont {Tokunaga},
  \citenamefont {Luo}, \citenamefont {Lin}, \citenamefont {Liu}, \citenamefont
  {Xiao}, \citenamefont {Miyake}, \citenamefont {Kohama}, \citenamefont {Lu},
  \citenamefont {Yang}, \citenamefont {Xia}, \citenamefont {Kindo},\ and\
  \citenamefont {Li}}]{ChenR2019}%
  \BibitemOpen
  \bibfield  {author} {\bibinfo {author} {\bibfnamefont {R.}~\bibnamefont
  {Chen}}, \bibinfo {author} {\bibfnamefont {J.~F.}\ \bibnamefont {Wang}},
  \bibinfo {author} {\bibfnamefont {Z.~W.}\ \bibnamefont {Ouyang}}, \bibinfo
  {author} {\bibfnamefont {M.}~\bibnamefont {Tokunaga}}, \bibinfo {author}
  {\bibfnamefont {A.~Y.}\ \bibnamefont {Luo}}, \bibinfo {author} {\bibfnamefont
  {L.}~\bibnamefont {Lin}}, \bibinfo {author} {\bibfnamefont {J.~M.}\
  \bibnamefont {Liu}}, \bibinfo {author} {\bibfnamefont {Y.}~\bibnamefont
  {Xiao}}, \bibinfo {author} {\bibfnamefont {A.}~\bibnamefont {Miyake}},
  \bibinfo {author} {\bibfnamefont {Y.}~\bibnamefont {Kohama}}, \bibinfo
  {author} {\bibfnamefont {C.~L.}\ \bibnamefont {Lu}}, \bibinfo {author}
  {\bibfnamefont {M.}~\bibnamefont {Yang}}, \bibinfo {author} {\bibfnamefont
  {Z.~C.}\ \bibnamefont {Xia}}, \bibinfo {author} {\bibfnamefont
  {K.}~\bibnamefont {Kindo}}, \ and\ \bibinfo {author} {\bibfnamefont
  {L.}~\bibnamefont {Li}},\ }\href {\doibase 10.1103/PhysRevB.100.140403}
  {\bibfield  {journal} {\bibinfo  {journal} {Phys. Rev. B}\ }\textbf {\bibinfo
  {volume} {100}},\ \bibinfo {pages} {140403} (\bibinfo {year}
  {2019})}\BibitemShut {NoStop}%
\bibitem [{\citenamefont {Wu}\ \emph {et~al.}(2020)\citenamefont {Wu},
  \citenamefont {Hsieh}, \citenamefont {Yen}, \citenamefont {Sun},
  \citenamefont {Kakarla}, \citenamefont {Her}, \citenamefont {Matsuda},
  \citenamefont {Chang}, \citenamefont {Lai}, \citenamefont {Gooch},
  \citenamefont {Deng}, \citenamefont {Webber}, \citenamefont {Lee},
  \citenamefont {Chou}, \citenamefont {Chu},\ and\ \citenamefont
  {Yang}}]{WuHC2020}%
  \BibitemOpen
  \bibfield  {author} {\bibinfo {author} {\bibfnamefont {H.~C.}\ \bibnamefont
  {Wu}}, \bibinfo {author} {\bibfnamefont {D.~J.}\ \bibnamefont {Hsieh}},
  \bibinfo {author} {\bibfnamefont {T.~W.}\ \bibnamefont {Yen}}, \bibinfo
  {author} {\bibfnamefont {P.~J.}\ \bibnamefont {Sun}}, \bibinfo {author}
  {\bibfnamefont {D.~C.}\ \bibnamefont {Kakarla}}, \bibinfo {author}
  {\bibfnamefont {J.~L.}\ \bibnamefont {Her}}, \bibinfo {author} {\bibfnamefont
  {Y.~H.}\ \bibnamefont {Matsuda}}, \bibinfo {author} {\bibfnamefont {C.~K.}\
  \bibnamefont {Chang}}, \bibinfo {author} {\bibfnamefont {Y.~C.}\ \bibnamefont
  {Lai}}, \bibinfo {author} {\bibfnamefont {M.}~\bibnamefont {Gooch}}, \bibinfo
  {author} {\bibfnamefont {L.~Z.}\ \bibnamefont {Deng}}, \bibinfo {author}
  {\bibfnamefont {K.~G.}\ \bibnamefont {Webber}}, \bibinfo {author}
  {\bibfnamefont {C.~A.}\ \bibnamefont {Lee}}, \bibinfo {author} {\bibfnamefont
  {M.~M.~C.}\ \bibnamefont {Chou}}, \bibinfo {author} {\bibfnamefont {C.~W.}\
  \bibnamefont {Chu}}, \ and\ \bibinfo {author} {\bibfnamefont {H.~D.}\
  \bibnamefont {Yang}},\ }\href {\doibase 10.1103/PhysRevB.102.075130}
  {\bibfield  {journal} {\bibinfo  {journal} {Phys. Rev. B}\ }\textbf {\bibinfo
  {volume} {102}},\ \bibinfo {pages} {075130} (\bibinfo {year}
  {2020})}\BibitemShut {NoStop}%
\bibitem [{\citenamefont {Guo}\ \emph {et~al.}(2015)\citenamefont {Guo},
  \citenamefont {Chemelewski}, \citenamefont {Mabayoje}, \citenamefont {Xiao},
  \citenamefont {Zhang},\ and\ \citenamefont {Mullins}}]{GuoW2015}%
  \BibitemOpen
  \bibfield  {author} {\bibinfo {author} {\bibfnamefont {W.}~\bibnamefont
  {Guo}}, \bibinfo {author} {\bibfnamefont {W.~D.}\ \bibnamefont
  {Chemelewski}}, \bibinfo {author} {\bibfnamefont {O.}~\bibnamefont
  {Mabayoje}}, \bibinfo {author} {\bibfnamefont {P.}~\bibnamefont {Xiao}},
  \bibinfo {author} {\bibfnamefont {Y.}~\bibnamefont {Zhang}}, \ and\ \bibinfo
  {author} {\bibfnamefont {C.~B.}\ \bibnamefont {Mullins}},\ }\href {\doibase
  10.1021/acs.jpcc.5b07219} {\bibfield  {journal} {\bibinfo  {journal} {The
  Journal of Physical Chemistry C}\ }\textbf {\bibinfo {volume} {119}},\
  \bibinfo {pages} {27220} (\bibinfo {year} {2015})},\ \Eprint
  {http://arxiv.org/abs/https://doi.org/10.1021/acs.jpcc.5b07219}
  {https://doi.org/10.1021/acs.jpcc.5b07219} \BibitemShut {NoStop}%
\bibitem [{\citenamefont {Yan}\ \emph {et~al.}(2015)\citenamefont {Yan},
  \citenamefont {Li}, \citenamefont {Newhouse}, \citenamefont {Yu},
  \citenamefont {Persson}, \citenamefont {Gregoire},\ and\ \citenamefont
  {Neaton}}]{YanQ2015}%
  \BibitemOpen
  \bibfield  {author} {\bibinfo {author} {\bibfnamefont {Q.}~\bibnamefont
  {Yan}}, \bibinfo {author} {\bibfnamefont {G.}~\bibnamefont {Li}}, \bibinfo
  {author} {\bibfnamefont {P.~F.}\ \bibnamefont {Newhouse}}, \bibinfo {author}
  {\bibfnamefont {J.}~\bibnamefont {Yu}}, \bibinfo {author} {\bibfnamefont
  {K.~A.}\ \bibnamefont {Persson}}, \bibinfo {author} {\bibfnamefont {J.~M.}\
  \bibnamefont {Gregoire}}, \ and\ \bibinfo {author} {\bibfnamefont {J.~B.}\
  \bibnamefont {Neaton}},\ }\href@noop {} {\bibfield  {journal} {\bibinfo
  {journal} {Advanced Energy Materials}\ }\textbf {\bibinfo {volume} {5}},\
  \bibinfo {pages} {1401840} (\bibinfo {year} {2015})}\BibitemShut {NoStop}%
\bibitem [{\citenamefont {woo Kim}\ \emph {et~al.}(2017)\citenamefont {woo
  Kim}, \citenamefont {Joshi}, \citenamefont {Yoon}, \citenamefont {Ohm},
  \citenamefont {Kim}, \citenamefont {Al-Deyab},\ and\ \citenamefont
  {Yoon}}]{KimMW2017}%
  \BibitemOpen
  \bibfield  {author} {\bibinfo {author} {\bibfnamefont {M.}~\bibnamefont {woo
  Kim}}, \bibinfo {author} {\bibfnamefont {B.}~\bibnamefont {Joshi}}, \bibinfo
  {author} {\bibfnamefont {H.}~\bibnamefont {Yoon}}, \bibinfo {author}
  {\bibfnamefont {T.~Y.}\ \bibnamefont {Ohm}}, \bibinfo {author} {\bibfnamefont
  {K.}~\bibnamefont {Kim}}, \bibinfo {author} {\bibfnamefont {S.~S.}\
  \bibnamefont {Al-Deyab}}, \ and\ \bibinfo {author} {\bibfnamefont {S.~S.}\
  \bibnamefont {Yoon}},\ }\href {\doibase
  https://doi.org/10.1016/j.jallcom.2017.02.302} {\bibfield  {journal}
  {\bibinfo  {journal} {Journal of Alloys and Compounds}\ }\textbf {\bibinfo
  {volume} {708}},\ \bibinfo {pages} {444} (\bibinfo {year}
  {2017})}\BibitemShut {NoStop}%
\bibitem [{\citenamefont {Camargo}\ \emph {et~al.}(2020)\citenamefont
  {Camargo}, \citenamefont {Lucilha}, \citenamefont {Gomes}, \citenamefont
  {Liberatti}, \citenamefont {Andrello}, \citenamefont {da~Silva},\ and\
  \citenamefont {Dall'Antonia}}]{CamargoLP2020}%
  \BibitemOpen
  \bibfield  {author} {\bibinfo {author} {\bibfnamefont {L.~P.}\ \bibnamefont
  {Camargo}}, \bibinfo {author} {\bibfnamefont {A.~C.}\ \bibnamefont
  {Lucilha}}, \bibinfo {author} {\bibfnamefont {G.~A.~B.}\ \bibnamefont
  {Gomes}}, \bibinfo {author} {\bibfnamefont {V.~R.}\ \bibnamefont
  {Liberatti}}, \bibinfo {author} {\bibfnamefont {A.~C.}\ \bibnamefont
  {Andrello}}, \bibinfo {author} {\bibfnamefont {P.~R.~C.}\ \bibnamefont
  {da~Silva}}, \ and\ \bibinfo {author} {\bibfnamefont {L.~H.}\ \bibnamefont
  {Dall'Antonia}},\ }\href@noop {} {\bibfield  {journal} {\bibinfo  {journal}
  {J. Solid State Electrochem.}\ }\textbf {\bibinfo {volume} {24}},\ \bibinfo
  {pages} {1935} (\bibinfo {year} {2020})}\BibitemShut {NoStop}%
\bibitem [{\citenamefont {Song}\ \emph {et~al.}(2020)\citenamefont {Song},
  \citenamefont {Chemseddine}, \citenamefont {Ahmet}, \citenamefont
  {Bogdanoff}, \citenamefont {Friedrich}, \citenamefont {Abdi}, \citenamefont
  {Berglund},\ and\ \citenamefont {van~de Krol}}]{SongA2020}%
  \BibitemOpen
  \bibfield  {author} {\bibinfo {author} {\bibfnamefont {A.}~\bibnamefont
  {Song}}, \bibinfo {author} {\bibfnamefont {A.}~\bibnamefont {Chemseddine}},
  \bibinfo {author} {\bibfnamefont {I.~Y.}\ \bibnamefont {Ahmet}}, \bibinfo
  {author} {\bibfnamefont {P.}~\bibnamefont {Bogdanoff}}, \bibinfo {author}
  {\bibfnamefont {D.}~\bibnamefont {Friedrich}}, \bibinfo {author}
  {\bibfnamefont {F.~F.}\ \bibnamefont {Abdi}}, \bibinfo {author}
  {\bibfnamefont {S.~P.}\ \bibnamefont {Berglund}}, \ and\ \bibinfo {author}
  {\bibfnamefont {R.}~\bibnamefont {van~de Krol}},\ }\href {\doibase
  10.1021/acs.chemmater.9b04909} {\bibfield  {journal} {\bibinfo  {journal}
  {Chemistry of Materials}\ }\textbf {\bibinfo {volume} {32}},\ \bibinfo
  {pages} {2408} (\bibinfo {year} {2020})}\BibitemShut {NoStop}%
\bibitem [{\citenamefont {Calvo}\ and\ \citenamefont
  {Faggiani}(1975{\natexlab{b}})}]{CalvoC1975}%
  \BibitemOpen
  \bibfield  {author} {\bibinfo {author} {\bibfnamefont {C.}~\bibnamefont
  {Calvo}}\ and\ \bibinfo {author} {\bibfnamefont {R.}~\bibnamefont
  {Faggiani}},\ }\href {\doibase 10.1107/S0567740875003354} {\bibfield
  {journal} {\bibinfo  {journal} {Acta Crystallographica Section B}\ }\textbf
  {\bibinfo {volume} {31}},\ \bibinfo {pages} {603} (\bibinfo {year}
  {1975}{\natexlab{b}})}\BibitemShut {NoStop}%
\bibitem [{\citenamefont {Robinson}\ \emph {et~al.}(1987)\citenamefont
  {Robinson}, \citenamefont {Hughes},\ and\ \citenamefont
  {Malinconico}}]{RobinsonPD1987}%
  \BibitemOpen
  \bibfield  {author} {\bibinfo {author} {\bibfnamefont {P.~D.}\ \bibnamefont
  {Robinson}}, \bibinfo {author} {\bibfnamefont {J.~M.}\ \bibnamefont
  {Hughes}}, \ and\ \bibinfo {author} {\bibfnamefont {M.~L.}\ \bibnamefont
  {Malinconico}},\ }\href@noop {} {\bibfield  {journal} {\bibinfo  {journal}
  {American Mineralogist}\ }\textbf {\bibinfo {volume} {72}},\ \bibinfo {pages}
  {297} (\bibinfo {year} {1987})}\BibitemShut {NoStop}%
\bibitem [{\citenamefont {Gitgeatpong}\ \emph {et~al.}(2015)\citenamefont
  {Gitgeatpong}, \citenamefont {Zhao}, \citenamefont {Avdeev}, \citenamefont
  {Piltz}, \citenamefont {Sato},\ and\ \citenamefont
  {Matan}}]{Gitgeatpong2015}%
  \BibitemOpen
  \bibfield  {author} {\bibinfo {author} {\bibfnamefont {G.}~\bibnamefont
  {Gitgeatpong}}, \bibinfo {author} {\bibfnamefont {Y.}~\bibnamefont {Zhao}},
  \bibinfo {author} {\bibfnamefont {M.}~\bibnamefont {Avdeev}}, \bibinfo
  {author} {\bibfnamefont {R.~O.}\ \bibnamefont {Piltz}}, \bibinfo {author}
  {\bibfnamefont {T.~J.}\ \bibnamefont {Sato}}, \ and\ \bibinfo {author}
  {\bibfnamefont {K.}~\bibnamefont {Matan}},\ }\href {\doibase
  10.1103/PhysRevB.92.024423} {\bibfield  {journal} {\bibinfo  {journal} {Phys.
  Rev. B}\ }\textbf {\bibinfo {volume} {92}},\ \bibinfo {pages} {024423}
  (\bibinfo {year} {2015})}\BibitemShut {NoStop}%
\bibitem [{\citenamefont {Banerjee}\ \emph {et~al.}(2016)\citenamefont
  {Banerjee}, \citenamefont {Sannigrahi}, \citenamefont {Bhowal}, \citenamefont
  {Dasgupta}, \citenamefont {Majumdar}, \citenamefont {Walker}, \citenamefont
  {Bhattacharyya},\ and\ \citenamefont {Adroja}}]{BanerjeeA2016}%
  \BibitemOpen
  \bibfield  {author} {\bibinfo {author} {\bibfnamefont {A.}~\bibnamefont
  {Banerjee}}, \bibinfo {author} {\bibfnamefont {J.}~\bibnamefont
  {Sannigrahi}}, \bibinfo {author} {\bibfnamefont {S.}~\bibnamefont {Bhowal}},
  \bibinfo {author} {\bibfnamefont {I.}~\bibnamefont {Dasgupta}}, \bibinfo
  {author} {\bibfnamefont {S.}~\bibnamefont {Majumdar}}, \bibinfo {author}
  {\bibfnamefont {H.~C.}\ \bibnamefont {Walker}}, \bibinfo {author}
  {\bibfnamefont {A.}~\bibnamefont {Bhattacharyya}}, \ and\ \bibinfo {author}
  {\bibfnamefont {D.~T.}\ \bibnamefont {Adroja}},\ }\href {\doibase
  10.1103/PhysRevB.94.144426} {\bibfield  {journal} {\bibinfo  {journal} {Phys.
  Rev. B}\ }\textbf {\bibinfo {volume} {94}},\ \bibinfo {pages} {144426}
  (\bibinfo {year} {2016})}\BibitemShut {NoStop}%
\bibitem [{\citenamefont {Chattopadhyay}\ \emph {et~al.}(2017)\citenamefont
  {Chattopadhyay}, \citenamefont {Ahmed}, \citenamefont {Bandyopadhyay},
  \citenamefont {Singha},\ and\ \citenamefont {Mandal}}]{ChattopadhyayB2017}%
  \BibitemOpen
  \bibfield  {author} {\bibinfo {author} {\bibfnamefont {B.}~\bibnamefont
  {Chattopadhyay}}, \bibinfo {author} {\bibfnamefont {M.~A.}\ \bibnamefont
  {Ahmed}}, \bibinfo {author} {\bibfnamefont {S.}~\bibnamefont
  {Bandyopadhyay}}, \bibinfo {author} {\bibfnamefont {R.}~\bibnamefont
  {Singha}}, \ and\ \bibinfo {author} {\bibfnamefont {P.}~\bibnamefont
  {Mandal}},\ }\href {\doibase 10.1063/1.4977859} {\bibfield  {journal}
  {\bibinfo  {journal} {Journal of Applied Physics}\ }\textbf {\bibinfo
  {volume} {121}},\ \bibinfo {pages} {094103} (\bibinfo {year} {2017})},\
  \Eprint {http://arxiv.org/abs/https://doi.org/10.1063/1.4977859}
  {https://doi.org/10.1063/1.4977859} \BibitemShut {NoStop}%
\bibitem [{\citenamefont {Gitgeatpong}\ \emph
  {et~al.}(2017{\natexlab{a}})\citenamefont {Gitgeatpong}, \citenamefont
  {Suewattana}, \citenamefont {Zhang}, \citenamefont {Miyake}, \citenamefont
  {Tokunaga}, \citenamefont {Chanlert}, \citenamefont {Kurita}, \citenamefont
  {Tanaka}, \citenamefont {Sato}, \citenamefont {Zhao},\ and\ \citenamefont
  {Matan}}]{GitgeatpongG2017}%
  \BibitemOpen
  \bibfield  {author} {\bibinfo {author} {\bibfnamefont {G.}~\bibnamefont
  {Gitgeatpong}}, \bibinfo {author} {\bibfnamefont {M.}~\bibnamefont
  {Suewattana}}, \bibinfo {author} {\bibfnamefont {S.}~\bibnamefont {Zhang}},
  \bibinfo {author} {\bibfnamefont {A.}~\bibnamefont {Miyake}}, \bibinfo
  {author} {\bibfnamefont {M.}~\bibnamefont {Tokunaga}}, \bibinfo {author}
  {\bibfnamefont {P.}~\bibnamefont {Chanlert}}, \bibinfo {author}
  {\bibfnamefont {N.}~\bibnamefont {Kurita}}, \bibinfo {author} {\bibfnamefont
  {H.}~\bibnamefont {Tanaka}}, \bibinfo {author} {\bibfnamefont {T.~J.}\
  \bibnamefont {Sato}}, \bibinfo {author} {\bibfnamefont {Y.}~\bibnamefont
  {Zhao}}, \ and\ \bibinfo {author} {\bibfnamefont {K.}~\bibnamefont {Matan}},\
  }\href {\doibase 10.1103/PhysRevB.95.245119} {\bibfield  {journal} {\bibinfo
  {journal} {Phys. Rev. B}\ }\textbf {\bibinfo {volume} {95}},\ \bibinfo
  {pages} {245119} (\bibinfo {year} {2017}{\natexlab{a}})}\BibitemShut
  {NoStop}%
\bibitem [{\citenamefont {Shiomi}\ \emph {et~al.}(2017)\citenamefont {Shiomi},
  \citenamefont {Takashima}, \citenamefont {Okuyama}, \citenamefont
  {Gitgeatpong}, \citenamefont {Piyawongwatthana}, \citenamefont {Matan},
  \citenamefont {Sato},\ and\ \citenamefont {Saitoh}}]{ShiomiY2017}%
  \BibitemOpen
  \bibfield  {author} {\bibinfo {author} {\bibfnamefont {Y.}~\bibnamefont
  {Shiomi}}, \bibinfo {author} {\bibfnamefont {R.}~\bibnamefont {Takashima}},
  \bibinfo {author} {\bibfnamefont {D.}~\bibnamefont {Okuyama}}, \bibinfo
  {author} {\bibfnamefont {G.}~\bibnamefont {Gitgeatpong}}, \bibinfo {author}
  {\bibfnamefont {P.}~\bibnamefont {Piyawongwatthana}}, \bibinfo {author}
  {\bibfnamefont {K.}~\bibnamefont {Matan}}, \bibinfo {author} {\bibfnamefont
  {T.~J.}\ \bibnamefont {Sato}}, \ and\ \bibinfo {author} {\bibfnamefont
  {E.}~\bibnamefont {Saitoh}},\ }\href {\doibase 10.1103/PhysRevB.96.180414}
  {\bibfield  {journal} {\bibinfo  {journal} {Phys. Rev. B}\ }\textbf {\bibinfo
  {volume} {96}},\ \bibinfo {pages} {180414} (\bibinfo {year}
  {2017})}\BibitemShut {NoStop}%
\bibitem [{\citenamefont {Zhang}\ \emph {et~al.}(2017)\citenamefont {Zhang},
  \citenamefont {Wang}, \citenamefont {Ji}, \citenamefont {Guo}, \citenamefont
  {Xia}, \citenamefont {Lu},\ and\ \citenamefont {Zhu}}]{ZhangJT2017}%
  \BibitemOpen
  \bibfield  {author} {\bibinfo {author} {\bibfnamefont {J.~T.}\ \bibnamefont
  {Zhang}}, \bibinfo {author} {\bibfnamefont {J.~L.}\ \bibnamefont {Wang}},
  \bibinfo {author} {\bibfnamefont {C.}~\bibnamefont {Ji}}, \bibinfo {author}
  {\bibfnamefont {B.~X.}\ \bibnamefont {Guo}}, \bibinfo {author} {\bibfnamefont
  {W.~S.}\ \bibnamefont {Xia}}, \bibinfo {author} {\bibfnamefont {X.~M.}\
  \bibnamefont {Lu}}, \ and\ \bibinfo {author} {\bibfnamefont {J.~S.}\
  \bibnamefont {Zhu}},\ }\href {\doibase 10.1103/PhysRevB.96.165132} {\bibfield
   {journal} {\bibinfo  {journal} {Phys. Rev. B}\ }\textbf {\bibinfo {volume}
  {96}},\ \bibinfo {pages} {165132} (\bibinfo {year} {2017})}\BibitemShut
  {NoStop}%
\bibitem [{\citenamefont {Gitgeatpong}\ \emph
  {et~al.}(2017{\natexlab{b}})\citenamefont {Gitgeatpong}, \citenamefont
  {Zhao}, \citenamefont {Piyawongwatthana}, \citenamefont {Qiu}, \citenamefont
  {Harriger}, \citenamefont {Butch}, \citenamefont {Sato},\ and\ \citenamefont
  {Matan}}]{GitgeatpongG2017b}%
  \BibitemOpen
  \bibfield  {author} {\bibinfo {author} {\bibfnamefont {G.}~\bibnamefont
  {Gitgeatpong}}, \bibinfo {author} {\bibfnamefont {Y.}~\bibnamefont {Zhao}},
  \bibinfo {author} {\bibfnamefont {P.}~\bibnamefont {Piyawongwatthana}},
  \bibinfo {author} {\bibfnamefont {Y.}~\bibnamefont {Qiu}}, \bibinfo {author}
  {\bibfnamefont {L.~W.}\ \bibnamefont {Harriger}}, \bibinfo {author}
  {\bibfnamefont {N.~P.}\ \bibnamefont {Butch}}, \bibinfo {author}
  {\bibfnamefont {T.~J.}\ \bibnamefont {Sato}}, \ and\ \bibinfo {author}
  {\bibfnamefont {K.}~\bibnamefont {Matan}},\ }\href {\doibase
  10.1103/PhysRevLett.119.047201} {\bibfield  {journal} {\bibinfo  {journal}
  {Phys. Rev. Lett.}\ }\textbf {\bibinfo {volume} {119}},\ \bibinfo {pages}
  {047201} (\bibinfo {year} {2017}{\natexlab{b}})}\BibitemShut {NoStop}%
\bibitem [{\citenamefont {Wang}\ \emph {et~al.}(2018)\citenamefont {Wang},
  \citenamefont {Werner}, \citenamefont {Ottmann}, \citenamefont {Weis},
  \citenamefont {Abdel-Hafiez}, \citenamefont {Sannigrahi}, \citenamefont
  {Majumdar}, \citenamefont {Koo},\ and\ \citenamefont
  {Klingeler}}]{WangL2018}%
  \BibitemOpen
  \bibfield  {author} {\bibinfo {author} {\bibfnamefont {L.}~\bibnamefont
  {Wang}}, \bibinfo {author} {\bibfnamefont {J.}~\bibnamefont {Werner}},
  \bibinfo {author} {\bibfnamefont {A.}~\bibnamefont {Ottmann}}, \bibinfo
  {author} {\bibfnamefont {R.}~\bibnamefont {Weis}}, \bibinfo {author}
  {\bibfnamefont {M.}~\bibnamefont {Abdel-Hafiez}}, \bibinfo {author}
  {\bibfnamefont {J.}~\bibnamefont {Sannigrahi}}, \bibinfo {author}
  {\bibfnamefont {S.}~\bibnamefont {Majumdar}}, \bibinfo {author}
  {\bibfnamefont {C.}~\bibnamefont {Koo}}, \ and\ \bibinfo {author}
  {\bibfnamefont {R.}~\bibnamefont {Klingeler}},\ }\href {\doibase
  10.1088/1367-2630/aac9dc} {\bibfield  {journal} {\bibinfo  {journal} {New
  Journal of Physics}\ }\textbf {\bibinfo {volume} {20}},\ \bibinfo {pages}
  {063045} (\bibinfo {year} {2018})}\BibitemShut {NoStop}%
\bibitem [{\citenamefont {Sannigrahi}\ \emph {et~al.}(2019)\citenamefont
  {Sannigrahi}, \citenamefont {Adroja}, \citenamefont {Perry}, \citenamefont
  {Gutmann}, \citenamefont {Petricek},\ and\ \citenamefont
  {Khalyavin}}]{SannigrahiJ2019}%
  \BibitemOpen
  \bibfield  {author} {\bibinfo {author} {\bibfnamefont {J.}~\bibnamefont
  {Sannigrahi}}, \bibinfo {author} {\bibfnamefont {D.~T.}\ \bibnamefont
  {Adroja}}, \bibinfo {author} {\bibfnamefont {R.}~\bibnamefont {Perry}},
  \bibinfo {author} {\bibfnamefont {M.~J.}\ \bibnamefont {Gutmann}}, \bibinfo
  {author} {\bibfnamefont {V.}~\bibnamefont {Petricek}}, \ and\ \bibinfo
  {author} {\bibfnamefont {D.}~\bibnamefont {Khalyavin}},\ }\href {\doibase
  10.1103/PhysRevMaterials.3.113401} {\bibfield  {journal} {\bibinfo  {journal}
  {Phys. Rev. Materials}\ }\textbf {\bibinfo {volume} {3}},\ \bibinfo {pages}
  {113401} (\bibinfo {year} {2019})}\BibitemShut {NoStop}%
\bibitem [{\citenamefont {Murashova}\ \emph {et~al.}(1993)\citenamefont
  {Murashova}, \citenamefont {Velikodnyi},\ and\ \citenamefont
  {Zhuravlev}}]{MurashovaEV1993}%
  \BibitemOpen
  \bibfield  {author} {\bibinfo {author} {\bibfnamefont {E.~V.}\ \bibnamefont
  {Murashova}}, \bibinfo {author} {\bibfnamefont {Y.~A.}\ \bibnamefont
  {Velikodnyi}}, \ and\ \bibinfo {author} {\bibfnamefont {V.~D.}\ \bibnamefont
  {Zhuravlev}},\ }\href@noop {} {\bibfield  {journal} {\bibinfo  {journal}
  {Russ. J. Inorg. Chem.}\ }\textbf {\bibinfo {volume} {38}},\ \bibinfo {pages}
  {1446} (\bibinfo {year} {1993})}\BibitemShut {NoStop}%
\bibitem [{\citenamefont {Momma}\ and\ \citenamefont
  {Izumi}(2011)}]{MommaK2011}%
  \BibitemOpen
  \bibfield  {author} {\bibinfo {author} {\bibfnamefont {K.}~\bibnamefont
  {Momma}}\ and\ \bibinfo {author} {\bibfnamefont {F.}~\bibnamefont {Izumi}},\
  }\href {\doibase 10.1107/S0021889811038970} {\bibfield  {journal} {\bibinfo
  {journal} {Journal of Applied Crystallography}\ }\textbf {\bibinfo {volume}
  {44}},\ \bibinfo {pages} {1272} (\bibinfo {year} {2011})}\BibitemShut
  {NoStop}%
\bibitem [{\citenamefont {Zhuravlev}\ \emph {et~al.}(2018)\citenamefont
  {Zhuravlev}, \citenamefont {Tyutyunnik}, \citenamefont {Chufarov},
  \citenamefont {Lobachevskaya},\ and\ \citenamefont
  {Velikodnyi}}]{ZhuravlevV2018}%
  \BibitemOpen
  \bibfield  {author} {\bibinfo {author} {\bibfnamefont {V.~D.}\ \bibnamefont
  {Zhuravlev}}, \bibinfo {author} {\bibfnamefont {A.~P.}\ \bibnamefont
  {Tyutyunnik}}, \bibinfo {author} {\bibfnamefont {A.~Y.}\ \bibnamefont
  {Chufarov}}, \bibinfo {author} {\bibfnamefont {N.~I.}\ \bibnamefont
  {Lobachevskaya}}, \ and\ \bibinfo {author} {\bibfnamefont {A.~A.}\
  \bibnamefont {Velikodnyi}},\ }\href {\doibase 10.1017/S0885715618000441}
  {\bibfield  {journal} {\bibinfo  {journal} {Powder Diffraction}\ }\textbf
  {\bibinfo {volume} {33}},\ \bibinfo {pages} {216–224} (\bibinfo {year}
  {2018})}\BibitemShut {NoStop}%
\bibitem [{\citenamefont {Babaryk}\ \emph {et~al.}(2015)\citenamefont
  {Babaryk}, \citenamefont {Odynets}, \citenamefont {Khainakov}, \citenamefont
  {Garcia-Granda},\ and\ \citenamefont {Slobodyanik}}]{BabarykAA2015}%
  \BibitemOpen
  \bibfield  {author} {\bibinfo {author} {\bibfnamefont {A.~A.}\ \bibnamefont
  {Babaryk}}, \bibinfo {author} {\bibfnamefont {I.~V.}\ \bibnamefont
  {Odynets}}, \bibinfo {author} {\bibfnamefont {S.}~\bibnamefont {Khainakov}},
  \bibinfo {author} {\bibfnamefont {S.}~\bibnamefont {Garcia-Granda}}, \ and\
  \bibinfo {author} {\bibfnamefont {N.~S.}\ \bibnamefont {Slobodyanik}},\
  }\href@noop {} {\bibfield  {journal} {\bibinfo  {journal} {CrystEngComm}\
  }\textbf {\bibinfo {volume} {17}},\ \bibinfo {pages} {7772} (\bibinfo {year}
  {2015})}\BibitemShut {NoStop}%
\bibitem [{\citenamefont {Vogt}\ and\ \citenamefont
  {M{\"u}ller-Buschbaum}(1991)}]{Vogt1991}%
  \BibitemOpen
  \bibfield  {author} {\bibinfo {author} {\bibfnamefont {R.}~\bibnamefont
  {Vogt}}\ and\ \bibinfo {author} {\bibfnamefont {H.}~\bibnamefont
  {M{\"u}ller-Buschbaum}},\ }\href@noop {} {\bibfield  {journal} {\bibinfo
  {journal} {Zeitschrift f{\"u}r anorganische und allgemeine Chemie}\ }\textbf
  {\bibinfo {volume} {594}},\ \bibinfo {pages} {119} (\bibinfo {year}
  {1991})}\BibitemShut {NoStop}%
\bibitem [{\citenamefont {Avdeev}\ and\ \citenamefont
  {Hester}(2018)}]{Avdeev2018}%
  \BibitemOpen
  \bibfield  {author} {\bibinfo {author} {\bibfnamefont {M.}~\bibnamefont
  {Avdeev}}\ and\ \bibinfo {author} {\bibfnamefont {J.~R.}\ \bibnamefont
  {Hester}},\ }\href {\doibase 10.1107/S1600576718014048} {\bibfield  {journal}
  {\bibinfo  {journal} {J. Appl. Crystallogr.}\ }\textbf {\bibinfo {volume}
  {51}},\ \bibinfo {pages} {1597} (\bibinfo {year} {2018})}\BibitemShut
  {NoStop}%
\bibitem [{\citenamefont {Rodriguez-Carvajal}(1993)}]{fullprof1993}%
  \BibitemOpen
  \bibfield  {author} {\bibinfo {author} {\bibfnamefont {J.}~\bibnamefont
  {Rodriguez-Carvajal}},\ }\href {\doibase
  https://doi.org/10.1016/0921-4526(93)90108-I} {\bibfield  {journal} {\bibinfo
   {journal} {Physica B: Condensed Matter}\ }\textbf {\bibinfo {volume}
  {192}},\ \bibinfo {pages} {55} (\bibinfo {year} {1993})}\BibitemShut
  {NoStop}%
\bibitem [{MSA()}]{MSAS2019}%
  \BibitemOpen
  \href@noop {} {}\bibinfo {note} {\noindent
  http://www2.tagen.tohoku.ac.jp/lab/sato\_tj/magnetic-representations-and-magnetic-space-groups/}\BibitemShut
  {NoStop}%
\bibitem [{\citenamefont {Bouloux}\ \emph {et~al.}(1972)\citenamefont
  {Bouloux}, \citenamefont {Perez},\ and\ \citenamefont
  {Galy}}]{BoulouxJC1972}%
  \BibitemOpen
  \bibfield  {author} {\bibinfo {author} {\bibfnamefont {J.~C.}\ \bibnamefont
  {Bouloux}}, \bibinfo {author} {\bibfnamefont {G.}~\bibnamefont {Perez}}, \
  and\ \bibinfo {author} {\bibfnamefont {J.}~\bibnamefont {Galy}},\ }\href@noop
  {} {\bibfield  {journal} {\bibinfo  {journal} {Bull. Soc. fr. Mineral.
  Cristallogr.}\ }\textbf {\bibinfo {volume} {95}},\ \bibinfo {pages} {130}
  (\bibinfo {year} {1972})}\BibitemShut {NoStop}%
\bibitem [{\citenamefont {Guida}\ and\ \citenamefont
  {Zinn-Justin}(1998)}]{GuidaR1998}%
  \BibitemOpen
  \bibfield  {author} {\bibinfo {author} {\bibfnamefont {R.}~\bibnamefont
  {Guida}}\ and\ \bibinfo {author} {\bibfnamefont {J.}~\bibnamefont
  {Zinn-Justin}},\ }\href {\doibase 10.1088/0305-4470/31/40/006} {\bibfield
  {journal} {\bibinfo  {journal} {Journal of Physics A: Mathematical and
  General}\ }\textbf {\bibinfo {volume} {31}},\ \bibinfo {pages} {8103}
  (\bibinfo {year} {1998})}\BibitemShut {NoStop}%
\bibitem [{\citenamefont {Izyumov}\ and\ \citenamefont
  {Naish}(1979)}]{Izyumov1979}%
  \BibitemOpen
  \bibfield  {author} {\bibinfo {author} {\bibfnamefont {Y.}~\bibnamefont
  {Izyumov}}\ and\ \bibinfo {author} {\bibfnamefont {V.}~\bibnamefont
  {Naish}},\ }\href {\doibase https://doi.org/10.1016/0304-8853(79)90086-6}
  {\bibfield  {journal} {\bibinfo  {journal} {J. Magn. Magn. Mater.}\ }\textbf
  {\bibinfo {volume} {12}},\ \bibinfo {pages} {239 } (\bibinfo {year}
  {1979})}\BibitemShut {NoStop}%
\bibitem [{\citenamefont {Izyumov}\ \emph {et~al.}(1991)\citenamefont
  {Izyumov}, \citenamefont {Naish},\ and\ \citenamefont
  {Ozerov}}]{Izyumov1991}%
  \BibitemOpen
  \bibfield  {author} {\bibinfo {author} {\bibfnamefont {Y.~A.}\ \bibnamefont
  {Izyumov}}, \bibinfo {author} {\bibfnamefont {V.~E.}\ \bibnamefont {Naish}},
  \ and\ \bibinfo {author} {\bibfnamefont {R.~P.}\ \bibnamefont {Ozerov}},\
  }\href@noop {} {\emph {\bibinfo {title} {Neutron Diffraction of Magnetic
  Materials}}}\ (\bibinfo  {publisher} {Springer Berlin Heidelberg},\ \bibinfo
  {address} {Berlin, Heidelberg},\ \bibinfo {year} {1991})\BibitemShut
  {NoStop}%
\bibitem [{\citenamefont {Belov}\ \emph {et~al.}(1957)\citenamefont {Belov},
  \citenamefont {Neronova},\ and\ \citenamefont {Smirnova}}]{BelovNV1957}%
  \BibitemOpen
  \bibfield  {author} {\bibinfo {author} {\bibfnamefont {N.~V.}\ \bibnamefont
  {Belov}}, \bibinfo {author} {\bibfnamefont {N.~N.}\ \bibnamefont {Neronova}},
  \ and\ \bibinfo {author} {\bibfnamefont {T.~S.}\ \bibnamefont {Smirnova}},\
  }\href@noop {} {\bibfield  {journal} {\bibinfo  {journal} {Sov. Phys.
  Crystallogr.}\ }\textbf {\bibinfo {volume} {2}},\ \bibinfo {pages} {311}
  (\bibinfo {year} {1957})}\BibitemShut {NoStop}%
\bibitem [{\citenamefont {Miller}\ and\ \citenamefont
  {Love}(1967)}]{MillerSC1967}%
  \BibitemOpen
  \bibfield  {author} {\bibinfo {author} {\bibfnamefont {S.~C.}\ \bibnamefont
  {Miller}}\ and\ \bibinfo {author} {\bibfnamefont {W.~F.}\ \bibnamefont
  {Love}},\ }\href@noop {} {\emph {\bibinfo {title} {Tables of irreducible
  representations os space groups and co-representations of magnetic space
  groups}}}\ (\bibinfo  {publisher} {Pruett Press, Boulder Colorado},\ \bibinfo
  {year} {1967})\BibitemShut {NoStop}%
\bibitem [{\citenamefont {Ortega-San~Martin}\ \emph {et~al.}(2005)\citenamefont
  {Ortega-San~Martin}, \citenamefont {Chapman}, \citenamefont {Lezama},
  \citenamefont {Sánchez-Marcos}, \citenamefont {Rodríguez-Fernández},
  \citenamefont {Arriortua},\ and\ \citenamefont {Rojo}}]{MartinOS2005}%
  \BibitemOpen
  \bibfield  {author} {\bibinfo {author} {\bibfnamefont {L.}~\bibnamefont
  {Ortega-San~Martin}}, \bibinfo {author} {\bibfnamefont {J.~P.}\ \bibnamefont
  {Chapman}}, \bibinfo {author} {\bibfnamefont {L.}~\bibnamefont {Lezama}},
  \bibinfo {author} {\bibfnamefont {J.}~\bibnamefont {Sánchez-Marcos}},
  \bibinfo {author} {\bibfnamefont {J.}~\bibnamefont {Rodríguez-Fernández}},
  \bibinfo {author} {\bibfnamefont {M.~I.}\ \bibnamefont {Arriortua}}, \ and\
  \bibinfo {author} {\bibfnamefont {T.}~\bibnamefont {Rojo}},\ }\href {\doibase
  10.1039/B413341B} {\bibfield  {journal} {\bibinfo  {journal} {J. Mater.
  Chem.}\ }\textbf {\bibinfo {volume} {15}},\ \bibinfo {pages} {183} (\bibinfo
  {year} {2005})}\BibitemShut {NoStop}%
\bibitem [{\citenamefont {Larrabee}\ \emph {et~al.}(2008)\citenamefont
  {Larrabee}, \citenamefont {Chyun},\ and\ \citenamefont
  {Volwiler}}]{LarrabeeJA2008}%
  \BibitemOpen
  \bibfield  {author} {\bibinfo {author} {\bibfnamefont {J.~A.}\ \bibnamefont
  {Larrabee}}, \bibinfo {author} {\bibfnamefont {S.-A.}\ \bibnamefont {Chyun}},
  \ and\ \bibinfo {author} {\bibfnamefont {A.~S.}\ \bibnamefont {Volwiler}},\
  }\href@noop {} {\bibfield  {journal} {\bibinfo  {journal} {Inorganic
  Chemistry}\ }\textbf {\bibinfo {volume} {47}},\ \bibinfo {pages} {10499}
  (\bibinfo {year} {2008})}\BibitemShut {NoStop}%
\bibitem [{\citenamefont {Daumann}\ \emph {et~al.}(2013)\citenamefont
  {Daumann}, \citenamefont {Comba}, \citenamefont {Larrabee}, \citenamefont
  {Schenk}, \citenamefont {Stranger}, \citenamefont {Cavigliasso},\ and\
  \citenamefont {Gahan}}]{DaumannLJ2013}%
  \BibitemOpen
  \bibfield  {author} {\bibinfo {author} {\bibfnamefont {L.~J.}\ \bibnamefont
  {Daumann}}, \bibinfo {author} {\bibfnamefont {P.}~\bibnamefont {Comba}},
  \bibinfo {author} {\bibfnamefont {J.~A.}\ \bibnamefont {Larrabee}}, \bibinfo
  {author} {\bibfnamefont {G.}~\bibnamefont {Schenk}}, \bibinfo {author}
  {\bibfnamefont {R.}~\bibnamefont {Stranger}}, \bibinfo {author}
  {\bibfnamefont {G.}~\bibnamefont {Cavigliasso}}, \ and\ \bibinfo {author}
  {\bibfnamefont {L.~R.}\ \bibnamefont {Gahan}},\ }\href@noop {} {\bibfield
  {journal} {\bibinfo  {journal} {Inorganic Chemistry}\ }\textbf {\bibinfo
  {volume} {52}},\ \bibinfo {pages} {2029} (\bibinfo {year} {2013})},\ \bibinfo
  {note} {pMID: 23374019}\BibitemShut {NoStop}%
\bibitem [{\citenamefont {Dutta}\ \emph {et~al.}(2007)\citenamefont {Dutta},
  \citenamefont {Seehra}, \citenamefont {Thota},\ and\ \citenamefont
  {Kumar}}]{DuttaP2007}%
  \BibitemOpen
  \bibfield  {author} {\bibinfo {author} {\bibfnamefont {P.}~\bibnamefont
  {Dutta}}, \bibinfo {author} {\bibfnamefont {M.~S.}\ \bibnamefont {Seehra}},
  \bibinfo {author} {\bibfnamefont {S.}~\bibnamefont {Thota}}, \ and\ \bibinfo
  {author} {\bibfnamefont {J.}~\bibnamefont {Kumar}},\ }\href {\doibase
  10.1088/0953-8984/20/01/015218} {\bibfield  {journal} {\bibinfo  {journal}
  {Journal of Physics: Condensed Matter}\ }\textbf {\bibinfo {volume} {20}},\
  \bibinfo {pages} {015218} (\bibinfo {year} {2007})}\BibitemShut {NoStop}%
\bibitem [{\citenamefont {Yogi}\ and\ \citenamefont
  {Isobe}(2021)}]{YogiAK2021}%
  \BibitemOpen
  \bibfield  {author} {\bibinfo {author} {\bibfnamefont {A.}~\bibnamefont
  {Yogi}}\ and\ \bibinfo {author} {\bibfnamefont {M.}~\bibnamefont {Isobe}},\
  }\href@noop {} {\enquote {\bibinfo {title} {Magnetic phase diagram and
  quantum critical point in the spin-1/2 zigzag-chain antiferromagnet
  cacov$_2$o$_7$},}\ }\bibinfo {howpublished} {Abstracts for the 2021 annual
  meeting of the Physical Society of Japan} (\bibinfo {year} {2021}),\ \bibinfo
  {note} {15aC1-7}\BibitemShut {NoStop}%
\end{thebibliography}%


%merlin.mbs apsrev4-1.bst 2010-07-25 4.21a (PWD, AO, DPC) hacked
%Control: key (0)
%Control: author (8) initials jnrlst
%Control: editor formatted (1) identically to author
%Control: production of article title (-1) disabled
%Control: page (0) single
%Control: year (1) truncated
%Control: production of eprint (0) enabled
\begin{thebibliography}{0}%
\makeatletter
\providecommand \@ifxundefined [1]{%
 \@ifx{#1\undefined}
}%
\providecommand \@ifnum [1]{%
 \ifnum #1\expandafter \@firstoftwo
 \else \expandafter \@secondoftwo
 \fi
}%
\providecommand \@ifx [1]{%
 \ifx #1\expandafter \@firstoftwo
 \else \expandafter \@secondoftwo
 \fi
}%
\providecommand \natexlab [1]{#1}%
\providecommand \enquote  [1]{``#1''}%
\providecommand \bibnamefont  [1]{#1}%
\providecommand \bibfnamefont [1]{#1}%
\providecommand \citenamefont [1]{#1}%
\providecommand \href@noop [0]{\@secondoftwo}%
\providecommand \href [0]{\begingroup \@sanitize@url \@href}%
\providecommand \@href[1]{\@@startlink{#1}\@@href}%
\providecommand \@@href[1]{\endgroup#1\@@endlink}%
\providecommand \@sanitize@url [0]{\catcode `\\12\catcode `\$12\catcode
  `\&12\catcode `\#12\catcode `\^12\catcode `\_12\catcode `\%12\relax}%
\providecommand \@@startlink[1]{}%
\providecommand \@@endlink[0]{}%
\providecommand \url  [0]{\begingroup\@sanitize@url \@url }%
\providecommand \@url [1]{\endgroup\@href {#1}{\urlprefix }}%
\providecommand \urlprefix  [0]{URL }%
\providecommand \Eprint [0]{\href }%
\providecommand \doibase [0]{http://dx.doi.org/}%
\providecommand \selectlanguage [0]{\@gobble}%
\providecommand \bibinfo  [0]{\@secondoftwo}%
\providecommand \bibfield  [0]{\@secondoftwo}%
\providecommand \translation [1]{[#1]}%
\providecommand \BibitemOpen [0]{}%
\providecommand \bibitemStop [0]{}%
\providecommand \bibitemNoStop [0]{.\EOS\space}%
\providecommand \EOS [0]{\spacefactor3000\relax}%
\providecommand \BibitemShut  [1]{\csname bibitem#1\endcsname}%
\let\auto@bib@innerbib\@empty
%</preamble>
\end{thebibliography}%

\end{document}

\title{Formation of ferromagnetic spin-dimers in the double pyrovanadate CaCoV$_2$O$_7$}% Force line breaks with \\
%%\thanks{A footnote to the article title}%

\author{Ryo Murasaki}
\email{r.murasaki@dc.tohoku.ac.jp}
\author{Kazuhiro Nawa}
\author{Daisuke Okuyama}

\affiliation{%
  Institute of Multidisciplinary Research for Advanced Materials, Tohoku University,
  2-1-1 Katahira, Sendai 980-8577, Japan
}%

\author{Maxim Avdeev}
\affiliation{Australian Centre for Neutron Scattering, Australian Nuclear Science and Technology Organisation,
  Locked Bag 2001, Kirrawee, NSW 2232, Australia}
\affiliation{School of Chemistry, The University of Sydney, Sydney, NSW 2006, Australia}

\author{Taku J Sato}
\email{taku@tohoku.ac.jp}
\affiliation{%
  Institute of Multidisciplinary Research for Advanced Materials, Tohoku University,
  2-1-1 Katahira, Sendai 980-8577, Japan
}%

\date{\today}% It is always \today, today,
             %  but any date may be explicitly specified

\begin{abstract}
  Neutron powder diffraction patterns for all the measured temperatures are given in this supplemental information, together with the fractional coordinates obtained from the Rietveld analysis.
\end{abstract}
\maketitle

\begin{figure}[h]
  \includegraphics[scale=0.3, angle=-90]{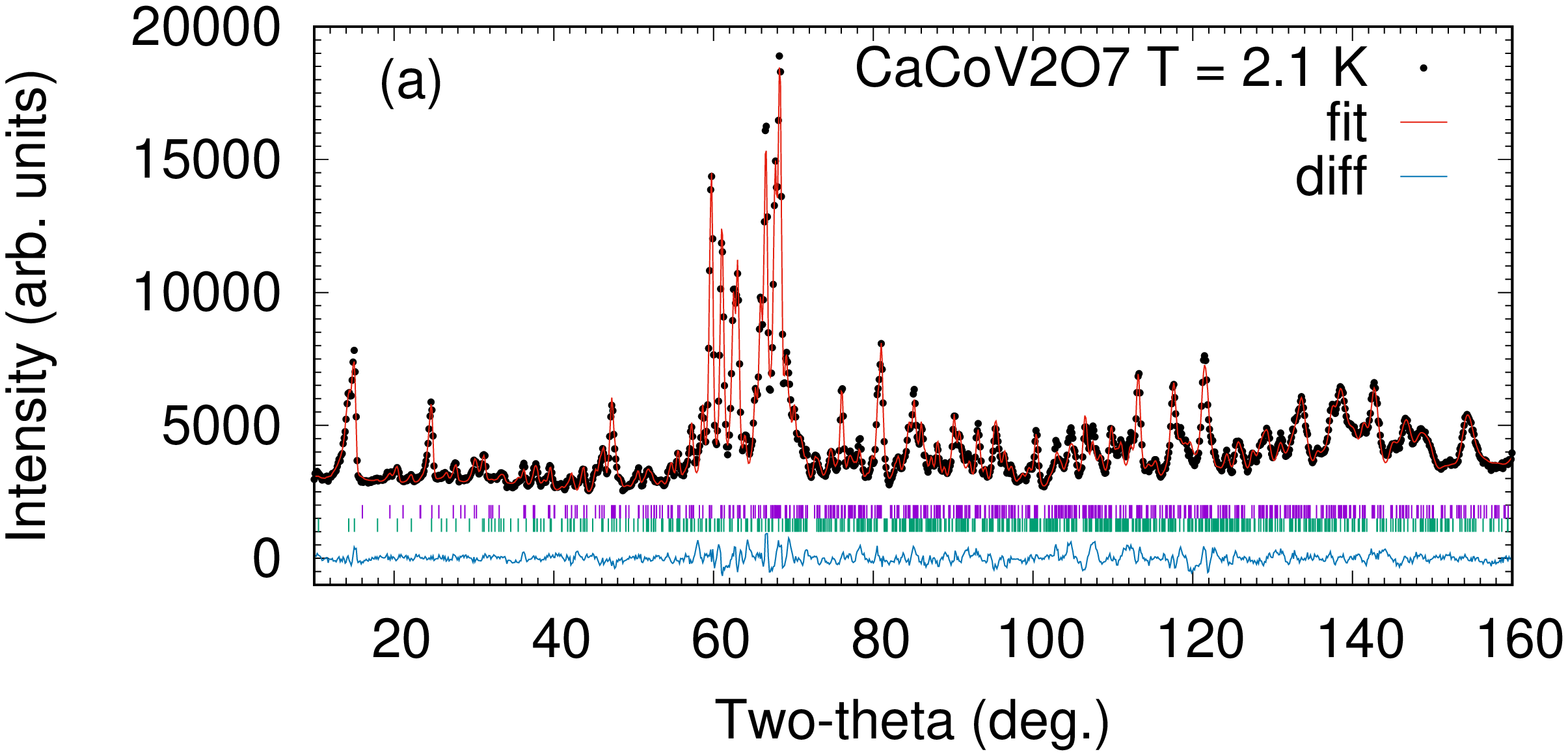}
  \includegraphics[scale=0.3, angle=-90]{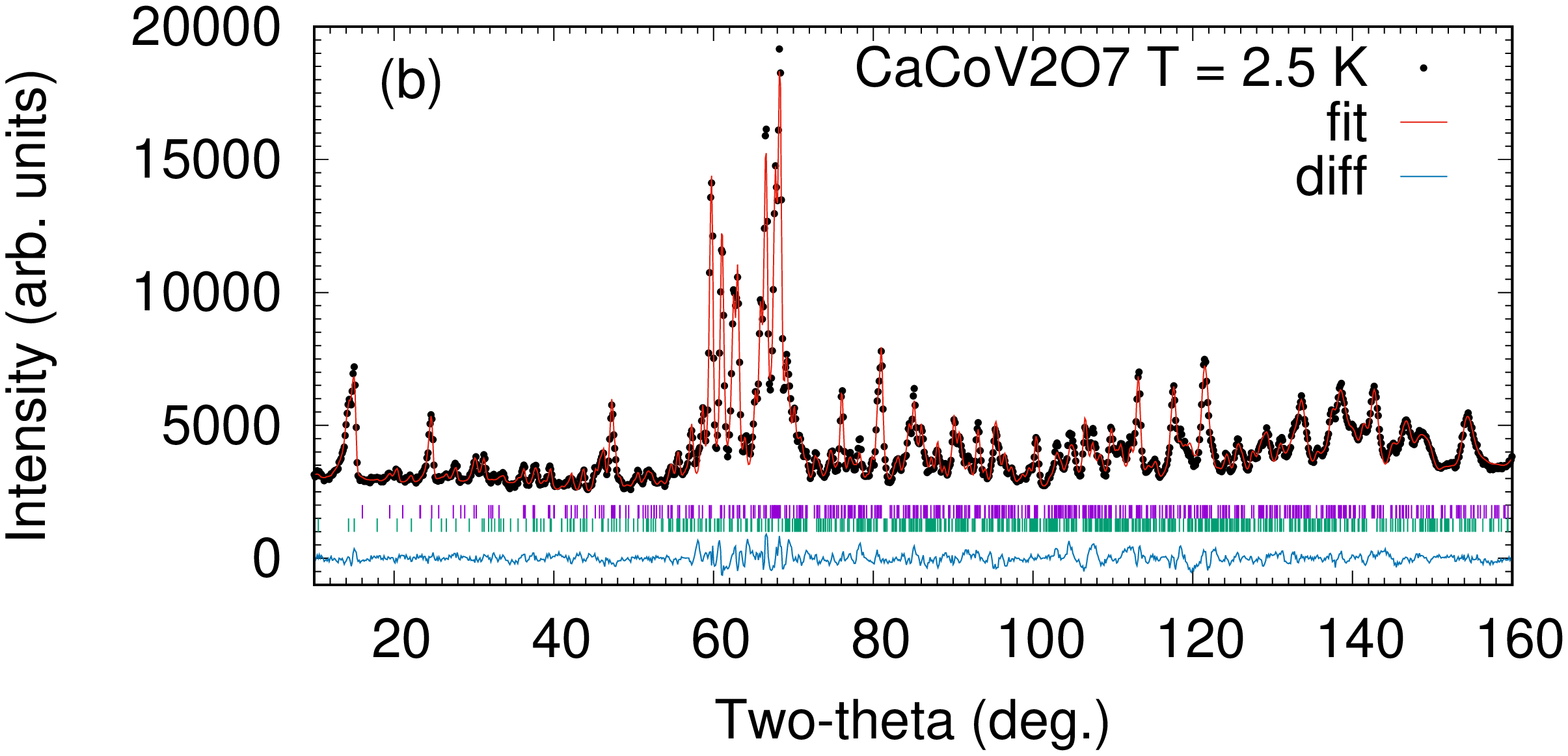}
  \includegraphics[scale=0.3, angle=-90]{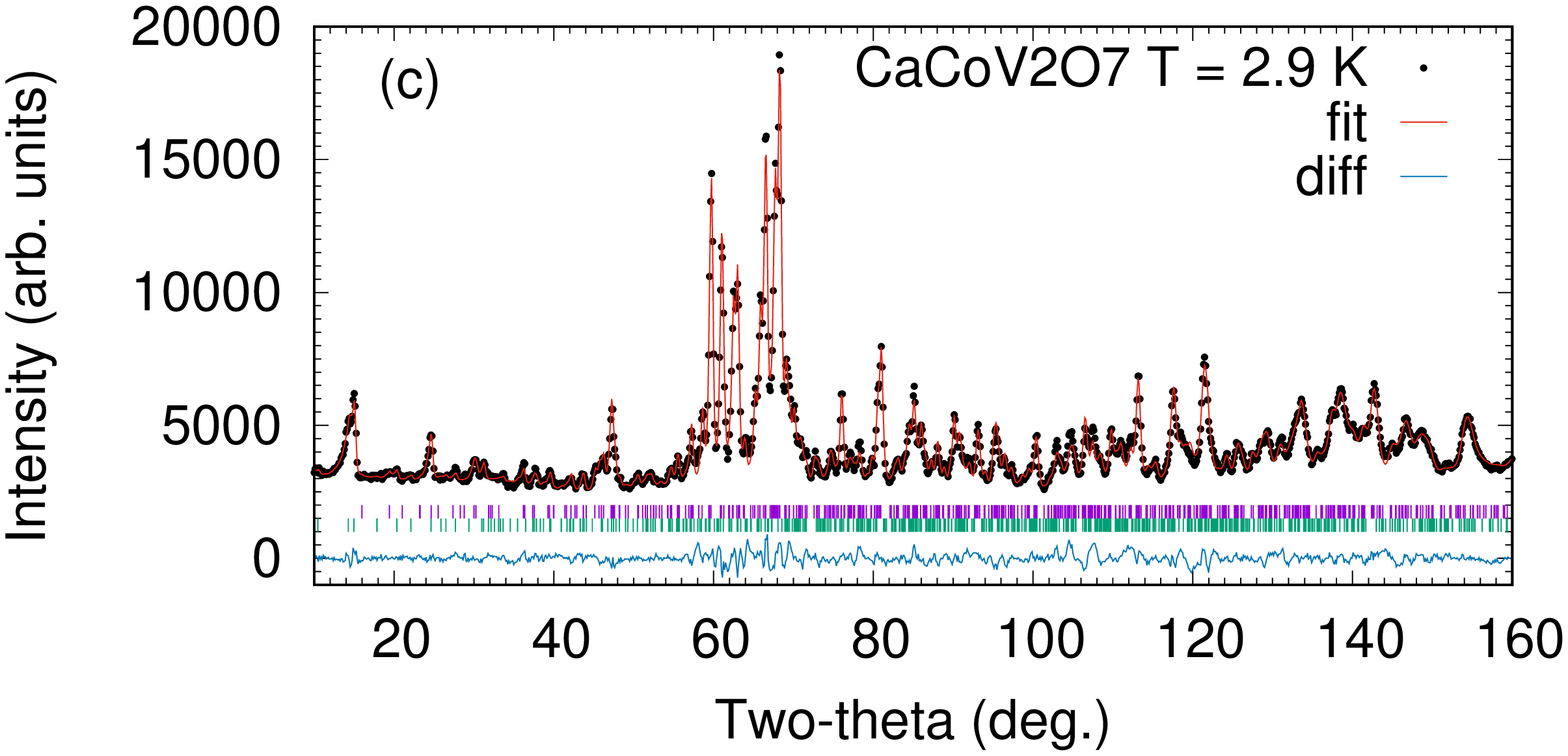}
  \includegraphics[scale=0.3, angle=-90]{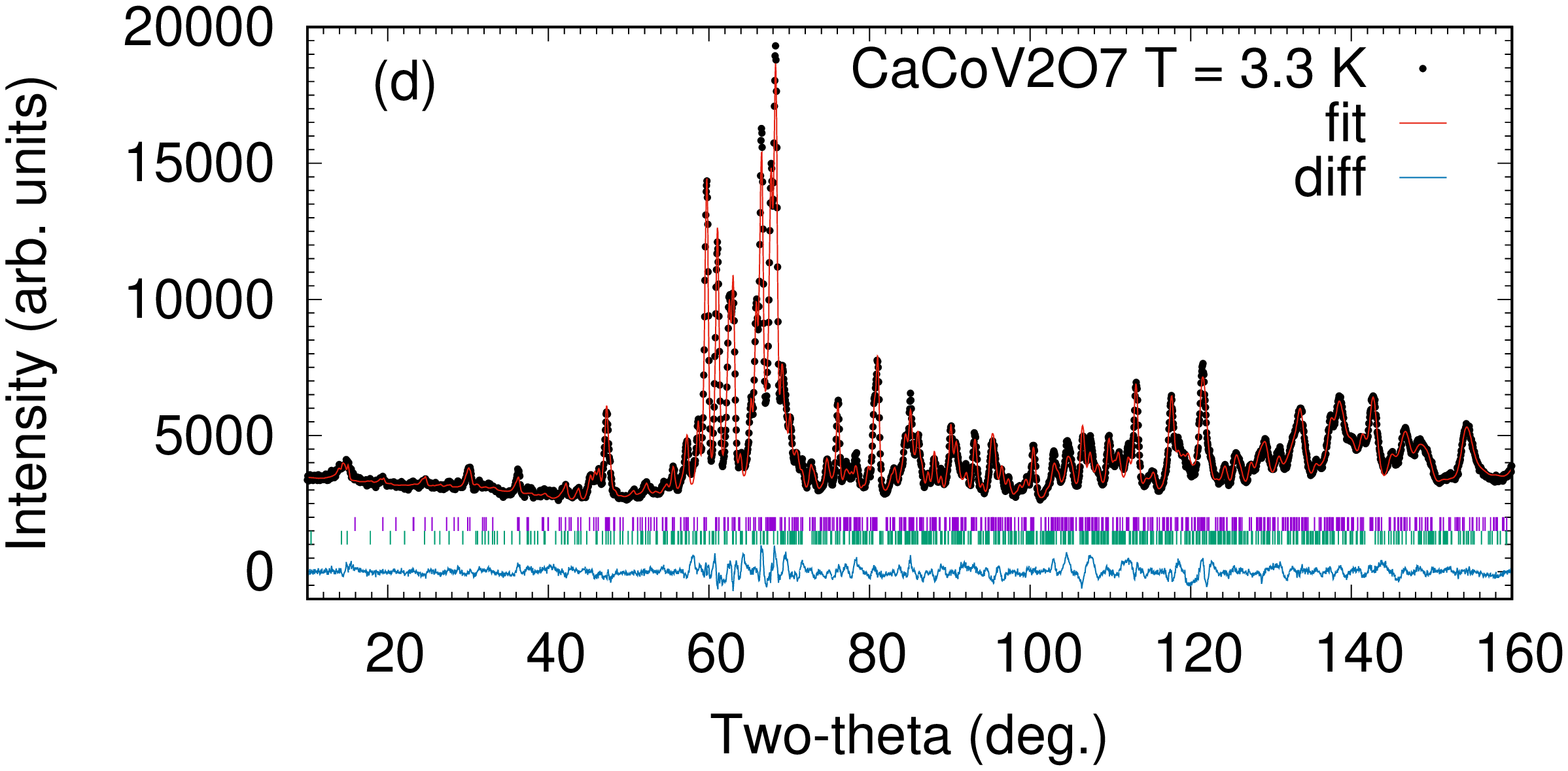}
  \includegraphics[scale=0.3, angle=-90]{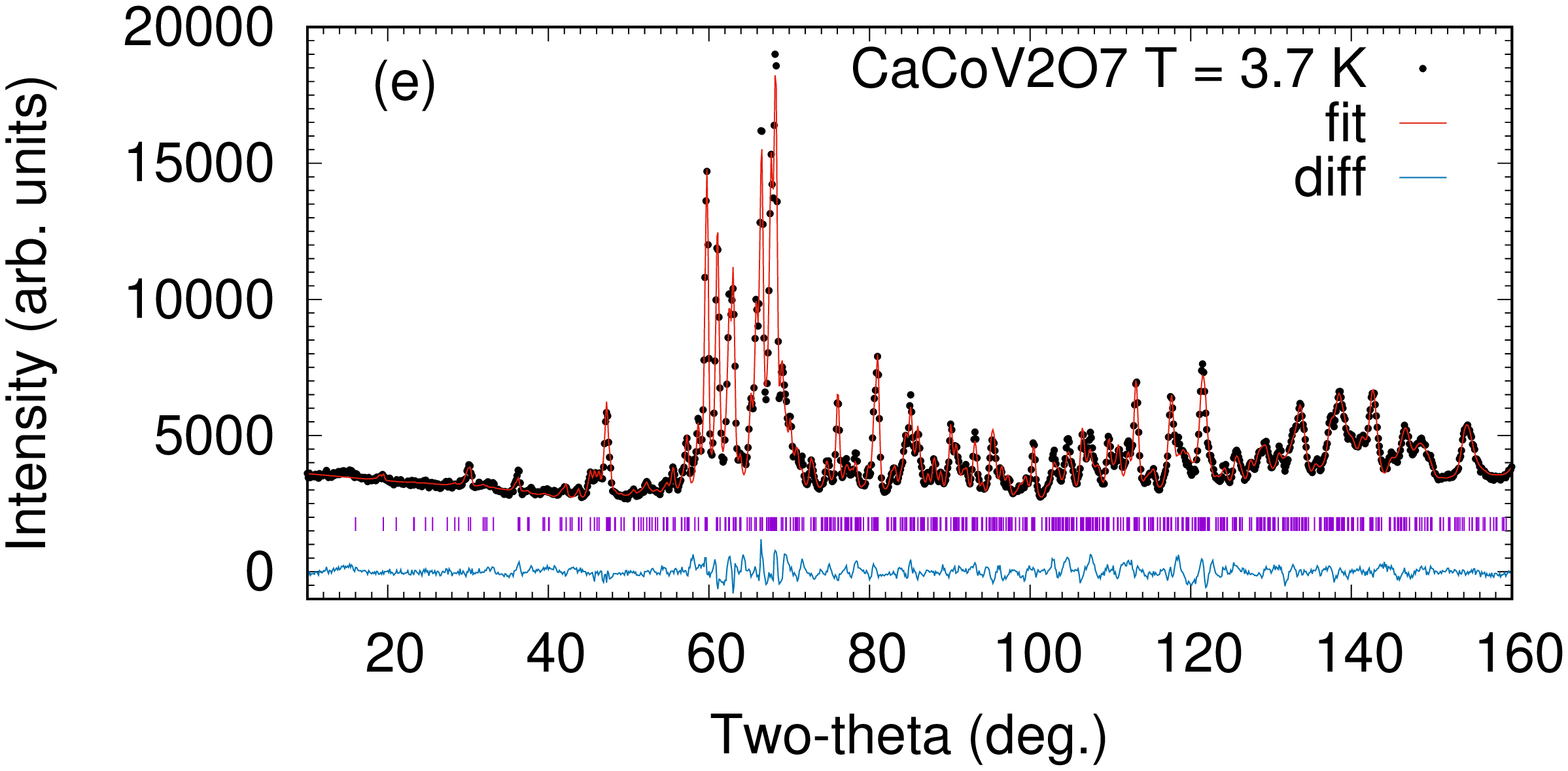}
  \includegraphics[scale=0.3, angle=-90]{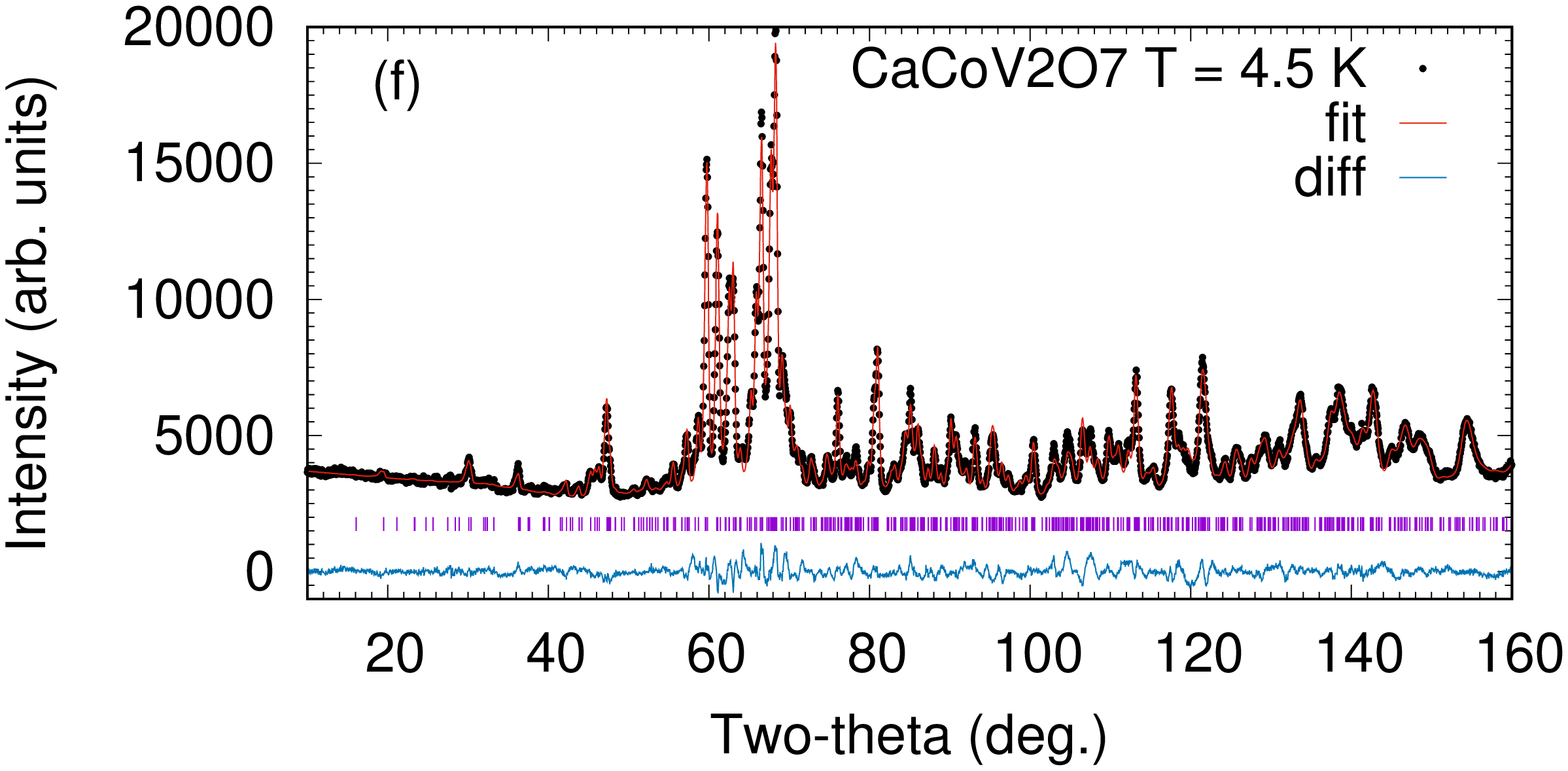}  
  \caption{(a-f) Neutron powder diffraction patterns of CaCoV$_2$O$_7$ measured at $T = 2.1$, 2.5, 2.9, 3.3, 3.7, and 4.5~K.
    The results of Rietveld fitting are also shown in the figure.
    Vertical lines indicate positions of nuclear (upper row), and magnetic (lower row) reflections, whereas the lines at the bottom in the plots stand for the difference between the observed and fitting results.
  }\label{sfig1}
\end{figure}

Neutron powder diffraction patterns at 2.1, 2.5, 2.9, 3.3, 3.7 and 4.5~K are shown in Fig.~\ref{sfig1}.
Fractional coordinate parameters obtained from the Rietveld analysis for all the temperatures (except those given in the manuscript) are shown in Tables~\ref{structure1.6K} to \ref{structure4.5K}.

\begin{table}[t]
\caption{Structure parameters of CaCoV$_2$O$_7$ at 1.6~K obtained in the present powder diffraction experiment.
The space group is $P2_1/c$, and the lattice parameter is $a$ = 6.7760(1) ~\AA, $b$ = 14.4450(2) ~\AA, $c$ = 11.1985(2)~\AA, $\beta$ = 100.1008(12)~\AA, .
The atomic coordinations are presented by fractional coordinates.
The equivalent isotropic displacement parameters $B_\mathrm{iso}$ are listed in a unit of \AA $^2$. Occupancy is fixed to 1 for all atoms.}
\label{structure1.6K}
\begin{center}
\begin{tabular}{lccccc}
\hline\hline
atom & site & $x$ & $y$ & $z$ & $B_\mathrm{iso}$ \\ \hline 
Ca1 & 4e & 0.6621(15) & 0.1308(8) & 0.0193(10) & 0.91(22) \\ 
Ca2 & 4e & 0.9347(13) & 0.0022(8) & 0.7517(10) & 0.91(22) \\ 
Co1 & 4e & $-$0.0131(26) & 0.2471(11) & 0.2664(16) & 0.25(0) \\ 
Co2 & 4e & 0.3492(24) & 0.1733(9) & 0.4288(14) & 0.25(0) \\ 
V1 & 4e & 0.4906(0) & 0.2055(0) & 0.7046(0) & 0.25(0) \\ 
V2 & 4e & 0.4215(0) & 0.0139(0) & 0.6872(0) & 0.25(0) \\ 
V3 & 4e & 0.1321(0) & $-$0.1348(0) & 0.5208(0) & 0.25(0) \\ 
V4 & 4e & 0.1582(0) & 0.3723(0) & 0.5144(0) & 0.25(0) \\ 
O1 & 4e & 0.5702(11) & 0.1066(6) & 0.7927(7) & 0.47(7) \\ 
O2 & 4e & 0.3107(13) & 0.2489(6) & 0.7644(8) & 0.47(7) \\ 
O3 & 4e & 0.3558(15) & 0.2993(6) & 0.5281(8) & 0.47(7) \\ 
O4 & 4e & 0.6977(14) & 0.2725(5) & 0.7025(8) & 0.47(7) \\ 
O5 & 4e & 0.4117(13) & 0.1215(6) & 0.5839(7) & 0.47(7) \\ 
O6 & 4e & 0.6188(12) & $-$0.0524(7) & 0.6767(8) & 0.47(7) \\ 
O7 & 4e & 0.2233(12) & $-$0.0309(7) & 0.5590(8) & 0.47(7) \\ 
O8 & 4e & 0.2985(11) & $-$0.0254(6) & 0.8011(7) & 0.47(7) \\ 
O9 & 4e & $-$0.0417(14) & $-$0.1557(6) & 0.6179(9) & 0.47(7) \\ 
O10 & 4e & 0.0152(13) & $-$0.1388(6) & 0.3775(9) & 0.47(7) \\ 
O11 & 4e & 0.3238(12) & $-$0.2160(6) & 0.5500(8) & 0.47(7) \\ 
O12 & 4e & 0.0199(14) & 0.3654(5) & 0.3725(8) & 0.47(7) \\ 
O13 & 4e & $-$0.0075(14) & 0.3536(6) & 0.6229(9) & 0.47(7) \\ 
O14 & 4e & 0.2591(10) & 0.4764(7) & 0.5244(7) & 0.47(7) \\ 
\hline\hline
\end{tabular}
\end{center}
\end{table}

\begin{table}[t]
\caption{Structure parameters of CaCoV$_2$O$_7$ at 2.1~K obtained in the present powder diffraction experiment.
The space group is $P2_1/c$, and the lattice parameter is $a$ = 6.7760(1) ~\AA, $b$ = 14.4447(4) ~\AA, $c$ = 11.1985(3)~\AA, $\beta$ = 100.1006(20)~\AA, .
The atomic coordinations are presented by fractional coordinates.
The equivalent isotropic displacement parameters $B_\mathrm{iso}$ are listed in a unit of \AA $^2$. Occupancy is fixed to 1 for all atoms.}
\label{structure2.1K}
\begin{center}
\begin{tabular}{lccccc}
\hline\hline
atom & site & $x$ & $y$ & $z$ & $B_\mathrm{iso}$ \\ \hline 
Ca1 & 4e & 0.6630(24) & 0.1305(14) & 0.0188(16) & 0.84(36) \\ 
Ca2 & 4e & 0.9353(20) & 0.0022(13) & 0.7521(15) & 0.84(36) \\ 
Co1 & 4e & $-$0.0113(42) & 0.2467(17) & 0.2652(27) & 0.25(0) \\ 
Co2 & 4e & 0.3486(40) & 0.1735(15) & 0.4281(23) & 0.25(0) \\ 
V1 & 4e & 0.4906(0) & 0.2055(0) & 0.7046(0) & 0.25(0) \\ 
V2 & 4e & 0.4215(0) & 0.0139(0) & 0.6872(0) & 0.25(0) \\ 
V3 & 4e & 0.1321(0) & $-$0.1348(0) & 0.5208(0) & 0.25(0) \\ 
V4 & 4e & 0.1582(0) & 0.3723(0) & 0.5144(0) & 0.25(0) \\ 
O1 & 4e & 0.5695(18) & 0.1063(10) & 0.7930(11) & 0.40(11) \\ 
O2 & 4e & 0.3111(21) & 0.2487(9) & 0.7646(12) & 0.40(11) \\ 
O3 & 4e & 0.3562(24) & 0.2996(10) & 0.5277(12) & 0.40(11) \\ 
O4 & 4e & 0.6979(22) & 0.2725(9) & 0.7024(13) & 0.40(11) \\ 
O5 & 4e & 0.4126(21) & 0.1215(10) & 0.5840(11) & 0.40(11) \\ 
O6 & 4e & 0.6196(20) & $-$0.0527(11) & 0.6771(13) & 0.40(11) \\ 
O7 & 4e & 0.2240(20) & $-$0.0313(11) & 0.5600(12) & 0.40(11) \\ 
O8 & 4e & 0.3003(19) & $-$0.0256(9) & 0.8011(12) & 0.40(11) \\ 
O9 & 4e & $-$0.0416(23) & $-$0.1564(9) & 0.6172(14) & 0.40(11) \\ 
O10 & 4e & 0.0158(22) & $-$0.1384(10) & 0.3775(14) & 0.40(11) \\ 
O11 & 4e & 0.3240(20) & $-$0.2162(9) & 0.5499(13) & 0.40(11) \\ 
O12 & 4e & 0.0182(23) & 0.3648(8) & 0.3716(13) & 0.40(11) \\ 
O13 & 4e & $-$0.0077(23) & 0.3541(9) & 0.6232(14) & 0.40(11) \\ 
O14 & 4e & 0.2582(16) & 0.4762(11) & 0.5240(11) & 0.40(11) \\ 
\hline\hline
\end{tabular}
\end{center}
\end{table}

\begin{table}[t]
\caption{Structure parameters of CaCoV$_2$O$_7$ at 2.5~K obtained in the present powder diffraction experiment.
The space group is $P2_1/c$, and the lattice parameter is $a$ = 6.7759(1) ~\AA, $b$ = 14.4451(4) ~\AA, $c$ = 11.1984(3)~\AA, $\beta$ = 100.1040(19)~\AA, .
The atomic coordinations are presented by fractional coordinates.
The equivalent isotropic displacement parameters $B_\mathrm{iso}$ are listed in a unit of \AA $^2$. Occupancy is fixed to 1 for all atoms.}
\label{structure2.5K}
\begin{center}
\begin{tabular}{lccccc}
\hline\hline
atom & site & $x$ & $y$ & $z$ & $B_\mathrm{iso}$ \\ \hline 
Ca1 & 4e & 0.6615(24) & 0.1310(14) & 0.0189(16) & 1.12(36) \\ 
Ca2 & 4e & 0.9373(21) & 0.0020(13) & 0.7532(16) & 1.12(36) \\ 
Co1 & 4e & $-$0.0150(43) & 0.2472(18) & 0.2651(27) & 0.25(0) \\ 
Co2 & 4e & 0.3502(40) & 0.1734(15) & 0.4294(23) & 0.25(0) \\ 
V1 & 4e & 0.4906(0) & 0.2055(0) & 0.7046(0) & 0.25(0) \\ 
V2 & 4e & 0.4215(0) & 0.0139(0) & 0.6872(0) & 0.25(0) \\ 
V3 & 4e & 0.1321(0) & $-$0.1348(0) & 0.5208(0) & 0.25(0) \\ 
V4 & 4e & 0.1582(0) & 0.3723(0) & 0.5144(0) & 0.25(0) \\ 
O1 & 4e & 0.5711(18) & 0.1062(10) & 0.7931(11) & 0.44(11) \\ 
O2 & 4e & 0.3103(20) & 0.2497(10) & 0.7643(12) & 0.44(11) \\ 
O3 & 4e & 0.3554(24) & 0.2998(10) & 0.5280(12) & 0.44(11) \\ 
O4 & 4e & 0.6984(21) & 0.2723(9) & 0.7019(13) & 0.44(11) \\ 
O5 & 4e & 0.4113(21) & 0.1216(10) & 0.5843(11) & 0.44(11) \\ 
O6 & 4e & 0.6189(19) & $-$0.0524(11) & 0.6766(12) & 0.44(11) \\ 
O7 & 4e & 0.2214(19) & $-$0.0301(11) & 0.5591(12) & 0.44(11) \\ 
O8 & 4e & 0.2995(18) & $-$0.0250(9) & 0.8019(11) & 0.44(11) \\ 
O9 & 4e & $-$0.0429(22) & $-$0.1557(9) & 0.6173(14) & 0.44(11) \\ 
O10 & 4e & 0.0163(21) & $-$0.1391(10) & 0.3767(14) & 0.44(11) \\ 
O11 & 4e & 0.3248(20) & $-$0.2161(9) & 0.5500(12) & 0.44(11) \\ 
O12 & 4e & 0.0195(23) & 0.3653(8) & 0.3721(12) & 0.44(11) \\ 
O13 & 4e & $-$0.0059(23) & 0.3538(9) & 0.6237(14) & 0.44(11) \\ 
O14 & 4e & 0.2598(16) & 0.4766(10) & 0.5241(11) & 0.44(11) \\ 
\hline\hline
\end{tabular}
\end{center}
\end{table}

\begin{table}[t]
\caption{Structure parameters of CaCoV$_2$O$_7$ at 2.9~K obtained in the present powder diffraction experiment.
The space group is $P2_1/c$, and the lattice parameter is $a$ = 6.7763(1) ~\AA, $b$ = 14.4453(4) ~\AA, $c$ = 11.1991(3)~\AA, $\beta$ = 100.1049(19)~\AA, .
The atomic coordinations are presented by fractional coordinates.
The equivalent isotropic displacement parameters $B_\mathrm{iso}$ are listed in a unit of \AA $^2$. Occupancy is fixed to 1 for all atoms.}
\label{structure2.9K}
\begin{center}
\begin{tabular}{lccccc}
\hline\hline
atom & site & $x$ & $y$ & $z$ & $B_\mathrm{iso}$ \\ \hline 
Ca1 & 4e & 0.6613(24) & 0.1308(14) & 0.0187(16) & 1.13(35) \\ 
Ca2 & 4e & 0.9367(20) & 0.0019(13) & 0.7518(15) & 1.13(35) \\ 
Co1 & 4e & $-$0.0222(44) & 0.2484(19) & 0.2688(27) & 0.25(0) \\ 
Co2 & 4e & 0.3484(41) & 0.1741(16) & 0.4279(24) & 0.25(0) \\ 
V1 & 4e & 0.4906(0) & 0.2055(0) & 0.7046(0) & 0.25(0) \\ 
V2 & 4e & 0.4215(0) & 0.0139(0) & 0.6872(0) & 0.25(0) \\ 
V3 & 4e & 0.1321(0) & $-$0.1348(0) & 0.5208(0) & 0.25(0) \\ 
V4 & 4e & 0.1582(0) & 0.3723(0) & 0.5144(0) & 0.25(0) \\ 
O1 & 4e & 0.5714(18) & 0.1060(9) & 0.7931(11) & 0.55(11) \\ 
O2 & 4e & 0.3080(20) & 0.2502(10) & 0.7642(12) & 0.55(11) \\ 
O3 & 4e & 0.3587(23) & 0.3001(10) & 0.5280(12) & 0.55(11) \\ 
O4 & 4e & 0.6966(21) & 0.2724(8) & 0.7020(13) & 0.55(11) \\ 
O5 & 4e & 0.4123(21) & 0.1217(10) & 0.5841(11) & 0.55(11) \\ 
O6 & 4e & 0.6198(19) & $-$0.0524(11) & 0.6771(12) & 0.55(11) \\ 
O7 & 4e & 0.2218(19) & $-$0.0297(11) & 0.5590(12) & 0.55(11) \\ 
O8 & 4e & 0.2986(18) & $-$0.0252(9) & 0.8020(11) & 0.55(11) \\ 
O9 & 4e & $-$0.0409(22) & $-$0.1556(9) & 0.6177(14) & 0.55(11) \\ 
O10 & 4e & 0.0172(21) & $-$0.1393(10) & 0.3780(14) & 0.55(11) \\ 
O11 & 4e & 0.3254(19) & $-$0.2164(9) & 0.5495(12) & 0.55(11) \\ 
O12 & 4e & 0.0207(22) & 0.3653(8) & 0.3725(12) & 0.55(11) \\ 
O13 & 4e & $-$0.0063(23) & 0.3537(9) & 0.6237(14) & 0.55(11) \\ 
O14 & 4e & 0.2608(15) & 0.4764(10) & 0.5243(11) & 0.55(11) \\ 
\hline\hline
\end{tabular}
\end{center}
\end{table}

\begin{table}[t]
\caption{Structure parameters of CaCoV$_2$O$_7$ at 3.3~K obtained in the present powder diffraction experiment.
The space group is $P2_1/c$, and the lattice parameter is $a$ = 6.7764(1) ~\AA, $b$ = 14.4463(3) ~\AA, $c$ = 11.1993(2)~\AA, $\beta$ = 100.1047(12)~\AA, .
The atomic coordinations are presented by fractional coordinates.
The equivalent isotropic displacement parameters $B_\mathrm{iso}$ are listed in a unit of \AA $^2$. Occupancy is fixed to 1 for all atoms.}
\label{atom3.3K}
\begin{center}
\begin{tabular}{lccccc}
\hline\hline
atom & site & $x$ & $y$ & $z$ & $B_\mathrm{iso}$ \\ \hline 
Ca1 & 4e & 0.6612(15) & 0.1310(9) & 0.0189(10) & 1.12(22) \\ 
Ca2 & 4e & 0.9358(12) & 0.0012(8) & 0.7519(10) & 1.12(22) \\ 
Co1 & 4e & $-$0.0178(32) & 0.2513(14) & 0.2702(17) & 0.25  \\ 
Co2 & 4e & 0.3470(29) & 0.1761(13) & 0.4255(16) & 0.25  \\ 
V1 & 4e & 0.4906  & 0.2055  & 0.7046  & 0.25  \\ 
V2 & 4e & 0.4215  & 0.0139  & 0.6872  & 0.25  \\ 
V3 & 4e & 0.1321  & $-$0.1348  & 0.5208  & 0.25  \\ 
V4 & 4e & 0.1582  & 0.3723  & 0.5144  & 0.25  \\ 
O1 & 4e & 0.5695(11) & 0.1059(6) & 0.7917(7) & 0.60(7) \\ 
O2 & 4e & 0.3096(13) & 0.2506(6) & 0.7637(8) & 0.60(7) \\ 
O3 & 4e & 0.3564(15) & 0.2991(6) & 0.5276(8) & 0.60(7) \\ 
O4 & 4e & 0.6969(13) & 0.2720(5) & 0.7019(8) & 0.60(7) \\ 
O5 & 4e & 0.4139(13) & 0.1218(6) & 0.5841(7) & 0.60(7) \\ 
O6 & 4e & 0.6187(12) & $-$0.0524(7) & 0.6768(8) & 0.60(7) \\ 
O7 & 4e & 0.2221(12) & $-$0.0291(7) & 0.5606(8) & 0.60(7) \\ 
O8 & 4e & 0.2998(11) & $-$0.0246(6) & 0.8005(7) & 0.60(7) \\ 
O9 & 4e & $-$0.0396(14) & $-$0.1551(6) & 0.6177(9) & 0.60(7) \\ 
O10 & 4e & 0.0199(13) & $-$0.1386(6) & 0.3774(9) & 0.60(7) \\ 
O11 & 4e & 0.3253(12) & $-$0.2158(6) & 0.5503(8) & 0.60(7) \\ 
O12 & 4e & 0.0180(14) & 0.3662(5) & 0.3727(8) & 0.60(7) \\ 
O13 & 4e & $-$0.0071(14) & 0.3540(6) & 0.6225(9) & 0.60(7) \\ 
O14 & 4e & 0.2601(10) & 0.4760(7) & 0.5252(7) & 0.60(7) \\ 
\hline\hline
\end{tabular}
\end{center}
\end{table}

\begin{table}[t]
\caption{Structure parameters of CaCoV$_2$O$_7$ at 3.7~K obtained in the present powder diffraction experiment.
The space group is $P2_1/c$, and the lattice parameter is $a$ = 6.7768(1) ~\AA, $b$ = 14.4477(4) ~\AA, $c$ = 11.2010(3)~\AA, $\beta$ = 100.1131(19)~\AA, .
The atomic coordinations are presented by fractional coordinates.
The equivalent isotropic displacement parameters $B_\mathrm{iso}$ are listed in a unit of \AA $^2$. Occupancy is fixed to 1 for all atoms.}
\label{atom3.7K}
\begin{center}
\begin{tabular}{lccccc}
\hline\hline
atom & site & $x$ & $y$ & $z$ & $B_\mathrm{iso}$ \\ \hline 
Ca1 & 4e & 0.6557(23) & 0.1323(13) & 0.0204(16) & 0.41(35) \\ 
Ca2 & 4e & 0.9396(19) & 0.0033(12) & 0.7509(15) & 0.41(35) \\ 
Co1 & 4e & $-$0.0198(50) & 0.2543(23) & 0.2730(28) & 0.25  \\ 
Co2 & 4e & 0.3549(47) & 0.1859(22) & 0.4252(28) & 0.25  \\ 
V1 & 4e & 0.4906  & 0.2055  & 0.7046  & 0.25  \\ 
V2 & 4e & 0.4215  & 0.0139  & 0.6872  & 0.25  \\ 
V3 & 4e & 0.1321  & $-$0.1348  & 0.5208  & 0.25  \\ 
V4 & 4e & 0.1582  & 0.3723  & 0.5144  & 0.25  \\ 
O1 & 4e & 0.5694(17) & 0.1050(10) & 0.7873(11) & 0.46(12) \\ 
O2 & 4e & 0.3136(20) & 0.2529(10) & 0.7639(12) & 0.46(12) \\ 
O3 & 4e & 0.3524(23) & 0.2894(11) & 0.5289(12) & 0.46(12) \\ 
O4 & 4e & 0.6975(21) & 0.2717(9) & 0.7023(13) & 0.46(12) \\ 
O5 & 4e & 0.4242(20) & 0.1231(11) & 0.5820(11) & 0.46(12) \\ 
O6 & 4e & 0.6217(21) & $-$0.0519(10) & 0.6739(12) & 0.46(12) \\ 
O7 & 4e & 0.2311(19) & $-$0.0173(12) & 0.5600(13) & 0.46(12) \\ 
O8 & 4e & 0.3010(18) & $-$0.0240(10) & 0.7986(11) & 0.46(12) \\ 
O9 & 4e & $-$0.0388(23) & $-$0.1552(9) & 0.6195(14) & 0.46(12) \\ 
O10 & 4e & 0.0236(22) & $-$0.1392(10) & 0.3753(14) & 0.46(12) \\ 
O11 & 4e & 0.3254(21) & $-$0.2060(13) & 0.5502(13) & 0.46(12) \\ 
O12 & 4e & 0.0166(23) & 0.3677(9) & 0.3746(12) & 0.46(12) \\ 
O13 & 4e & $-$0.0052(24) & 0.3551(9) & 0.6180(14) & 0.46(12) \\ 
O14 & 4e & 0.2613(17) & 0.4665(11) & 0.5216(12) & 0.46(12) \\ 
\hline\hline
\end{tabular}
\end{center}
\end{table}

\begin{table}[t]
\caption{Structure parameters of CaCoV$_2$O$_7$ at 4.5~K obtained in the present powder diffraction experiment.
The space group is $P2_1/c$, and the lattice parameter is $a$ = 6.7765(1) ~\AA, $b$ = 14.4468(3) ~\AA, $c$ = 11.1994(2)~\AA, $\beta$ = 100.1086(12)~\AA, .
The atomic coordinations are presented by fractional coordinates.
The equivalent isotropic displacement parameters $B_\mathrm{iso}$ are listed in a unit of \AA $^2$. Occupancy is fixed to 1 for all atoms.}
\label{structure4.5K}
\begin{center}
\begin{tabular}{lccccc}
\hline\hline
atom & site & $x$ & $y$ & $z$ & $B_\mathrm{iso}$ \\ \hline 
Ca1 & 4e & 0.6616(15) & 0.1321(8) & 0.0191(10) & 0.72(22) \\ 
Ca2 & 4e & 0.9340(12) & 0.0006(8) & 0.7514(9) & 0.72(22) \\ 
Co1 & 4e & $-$0.0195(32) & 0.2518(14) & 0.2745(17) & 0.25(0) \\ 
Co2 & 4e & 0.3437(30) & 0.1762(14) & 0.4257(17) & 0.25(0) \\ 
V1 & 4e & 0.4906(0) & 0.2055(0) & 0.7046(0) & 0.25(0) \\ 
V2 & 4e & 0.4215(0) & 0.0139(0) & 0.6872(0) & 0.25(0) \\ 
V3 & 4e & 0.1321(0) & $-$0.1348(0) & 0.5208(0) & 0.25(0) \\ 
V4 & 4e & 0.1582(0) & 0.3723(0) & 0.5144(0) & 0.25(0) \\ 
O1 & 4e & 0.5698(11) & 0.1054(6) & 0.7916(7) & 0.40(7) \\ 
O2 & 4e & 0.3102(13) & 0.2509(6) & 0.7648(8) & 0.40(7) \\ 
O3 & 4e & 0.3560(15) & 0.2991(6) & 0.5272(8) & 0.40(7) \\ 
O4 & 4e & 0.6970(14) & 0.2717(5) & 0.7023(8) & 0.40(7) \\ 
O5 & 4e & 0.4159(13) & 0.1223(6) & 0.5837(7) & 0.40(7) \\ 
O6 & 4e & 0.6195(12) & $-$0.0518(7) & 0.6759(8) & 0.40(7) \\ 
O7 & 4e & 0.2214(12) & $-$0.0289(7) & 0.5599(8) & 0.40(7) \\ 
O8 & 4e & 0.2976(11) & $-$0.0237(6) & 0.7998(7) & 0.40(7) \\ 
O9 & 4e & $-$0.0411(14) & $-$0.1560(6) & 0.6179(9) & 0.40(7) \\ 
O10 & 4e & 0.0170(13) & $-$0.1394(6) & 0.3765(9) & 0.40(7) \\ 
O11 & 4e & 0.3236(12) & $-$0.2154(6) & 0.5508(8) & 0.40(7) \\ 
O12 & 4e & 0.0176(14) & 0.3664(5) & 0.3721(8) & 0.40(7) \\ 
O13 & 4e & $-$0.0063(15) & 0.3539(6) & 0.6226(9) & 0.40(7) \\ 
O14 & 4e & 0.2605(10) & 0.4748(7) & 0.5248(7) & 0.40(7) \\ 
\hline\hline
\end{tabular}
\end{center}
\end{table}